# Quasi-universality in single-cell sequencing data

Luis Aparicio[1,2,+], Mykola Bordyuh[1,2,+], Andrew J. Blumberg[3], Raul Rabadan[1,2,#]

[1]Department of Systems Biology and [2]Department of Biomedical Informatics

Columbia University, New York, NY10032, US

[3]Department of Mathematics

University of Texas, Austin, US

[+]These authors contributed equally to this work

[#]Corresponding author: rr2579@cumc.columbia.edu

## ABSTRACT

The development of single-cell technologies provides the opportunity to identify new cellular states and reconstruct novel cell-to-cell relationships. Applications range from understanding the transcriptional and epigenetic processes involved in metazoan development to characterizing distinct cells types in heterogeneous populations like cancers or immune cells. However, analysis of the data is impeded by its unknown intrinsic biological and technical variability together with its sparseness; these factors complicate the identification of true biological signals amidst artifact and noise. Here we show that, across technologies, roughly 95% of the eigenvalues derived from each single-cell data set can be described by universal distributions predicted by Random Matrix Theory. Interestingly, 5% of the spectrum shows deviations from these distributions and present a phenomenon known as eigenvector localization, where information tightly concentrates in groups of cells. Some of the localized eigenvectors reflect underlying biological signal, and some are simply a consequence of the sparsity of single cell data; roughly 3% is artifactual. Based on the universal distributions and a technique for detecting sparsity induced localization, we present a strategy to identify the residual 2% of directions that encode biological information and thereby denoise single-cell data. We demonstrate the effectiveness of this approach by comparing with standard single-cell data analysis techniques in a variety of examples with marked cell populations.

## **INTRODUCTION**

Single-cell technologies offer the opportunity to identify previously unreported cell types and cellular states and explore the relationship between new and known cell states (*1-7*). However, there exist several significant biological and technical challenges that complicate the analysis. The first challenge relates to the lack of a complete quantitative understanding of the different sorts of noise. Almost identical cells have an intrinsic cell-to-cell variability and, within a cell, there are spatial and temporal fluctuations. Moreover, different technologies show biases at the level of detection, amplifying, and sequencing genomic material that significantly vary across different genomic loci. Estimating noise, and distinguishing between biological and technical sources, is paramount for any further analysis: without reliable estimates of noise it is difficult to distinguish states or identify potential variations of a single state. A second complicating factor for single-cell analysis is the sparsity of data associated to the very low amounts of genomic material amplified.

Several computational and statistical approaches have been designed to address some of these challenges (*4, 8-13*). For instance, imputation methods try to infer the "true" expression for missing values from the sample data by empirically modeling the underlying distributions, for instance, using negative binomial plus zero inflation (drop-out) for single-cell data.  These techniques usually assume that values are generated by the same distribution (identically independent distributed variables or i.i.d.).  However, we currently do not have predictive quantitative models of gene expression and so it is not clear what is the correct distribution or why the i.i.d. assumption should hold.  Given the lack of a quantitative microscopic description of cell transcription, we would ideally like to have a statistical description of the noise in single-cell data that does not rely on specific details of the underlying distributions of expression.

Historically, a similar problem occurred in the 1950s in nuclear physics, when the lack of quantitative models of complex nuclei precluded accurate predictions of their energy levels. However, simple theoretical models based on experimental data showed that some observables, such as the distance between two consecutive energy levels, followed universal distributions that could be derived from random Hermitian matrices (*14-16*). These distributions were subsequently identified in a variety of complex systems, as quantum versions of chaotic systems (*17*), and even in patterns of zeros of Riemann zeta functions (*18, 19*). Further work showed that many random matrix statistics present a universal behavior, akin to the central limit theorem, where specific details of the underlying distribution generating the entries of the matrix become irrelevant (*20, 21*). In the context of PCA problems with where the ratio of the rows and columns converges to a constant as both dimensions go to infinity, as arises in the single-cell setting, the study of the asymptotics of random matrices led to development of techniques for sparse PCA. These methods are intended to correct for bias in empirical eigenvalues and eigenvectors in PCA and enhance interpretability of the results (*22, 23*).

We propose here to apply these asymptotics to identify universal statistical features of noise that are insensitive to the specific details of a complex system (i.e., the cell and the single-cell measurement technologies). It has recently been shown that universality of the eigenvalues of random matrices depend only on the asymptotic behavior (subexponential) or the finiteness of the first few moments of the distribution generating the matrix, without requirement of being identically distributed random variables (*24-26*). These hypotheses hold in all distributions commonly used to describe single-cell data. Similar strong results have been observed in the distribution of eigenvectors: eigenvector of a random matrix show a phenomenon called de-localization, being distributed randomly in a high dimensional sphere. These universal characterizations provide the basis of a test: Deviations from universal eigenvalue distributions (*20, 27*) or the appearance of localized eigenvectors indicate the presence of a signal that can be further analyzed.

However, this test can be confounded by sparsity. Specifically, we show that the intrinsic sparsity of single-cell data can introduce deviations from the universal random matrix eigenvalue distribution. Nonetheless, we observe that these deviations can be easily identified by the presence of localized eigenvectors that survive permutation of the data. We combine the universal random matrix statistics with corrections for sparsity to identify the biological signal in single-cell data. By studying a variety single-cell transcriptomic experiments, we show that the spectrum of a normalized Wishart matrix generated from the data follows a Marchenko-Pastur (MP) distribution with a small fraction of outliers eigenvalues and localized eigenvectors. Thus, we can use the associated edge statistics (Tracy-Widom) and localized eigenvectors to generate a low rank approximation of the data, increasing the power for identifying potential interesting biological signals. Our direct approach is substantially simpler and more efficient than sparse PCA (*22, 23*), explicitly handles sparsity of the data matrix (without any assumptions about the distribution of missing elements), and leads directly to an estimate of the rank of the underlying signal. We show that this procedure is better able to capture marked single-cell clusters than alternative techniques, across a variety of data sets.

# RESULTS

**Quasi-universality of single-cell sequencing data**.

A standard technique for studying high dimensional data is dimensionality reduction via Principal Component Analysis (PCA) or versions thereof. PCA estimates the sample covariance and can be used to generate a low dimensional representation by projecting into the eigenvectors associated with the eigenvalues capturing most of the variance (principal components). We observed that the distribution of separation between the square root of two consecutive eigenvalues of the sample covariance matrix in different single-cell RNASeq experiments (*28-33*) resembles the Wigner surmise distribution conjectured by Wigner in 1955 (*16*) in the study of the difference between resonant peaks in slow neutron scattering (Figure 1A, and Supplementary Figure 1). This observation prompted us to investigate the connection between Random Matrix Theory (RMT) and the spectra of single-cell data, guided by the hypothesis that departures from random matrix distributions indicate interesting potential biological signals. Figure 1B shows an example, in red the non-parametric Marchenko-Pastur distribution (MP) of density of eigenvalues, the associated RMT distribution, that fits well most of the eigenvalues (see Methods). The same results can be observed across many other single-cell datasets (Supplementary Figure 2, Methods). Deviations from RMT can also be found by analyzing the larger eigenvalues in relations to the expected Tracy-Widom distribution (TW). We observed that across single-cell datasets this deviations amount to 5% of eigenvalues (Figure 1C).

We further investigated the potential leading causes explaining these deviations. After randomizing the data by shuffling the cell expression values in each gene independently to erase any potential signal in the sample covariance matrix, we found that only ~2% of eigenvalues could be associated with potential biological signal (Figure 1C).

**Sparsity Induced Eigenvector Localization**

One of the key features of single-cell data is its sparsity. We investigated if sparsity could induce deviations from universality. By introducing zeros in a random matrix with entries generated with Gaussian or Poisson distributions, we observed that the mean and the median of the highest eigenvalue have significant deviations from the RMT Tracy-Widom distribution (Figure 2A, Supplementary Figure 3). A similar phenomenon has been reported in the case of Wigner matrices (Hermitian matrices) in the context of Sparse Random Matrix Ensembles, a generalization of RMT with random matrices with a significant fraction of zero entries (*24, 34-36*). Numerical experiments strongly suggest that if the number of non-zero values per column is larger than an undetermined constant, there is phase transition where the density distribution of eigenvalues deviates from MP and that eigenvectors become localized. The deviations from the non-sparse RMT distributions have been associated to the phenomenon of Anderson localization: whereas in non-sparse random matrices eigenvector are randomly distributed in a sphere (Haar measure in orthogonal groups), in sparse matrices eigenvectors are localized along particular directions. We observe that this transition also occurs in the Wishart/covariance ensemble at some value of non-zero entries per row. In the case of medium sparsity (fraction of non-zero values $p > N^{-1/3}$, where N is the size of the matrix) it was recently reported (*37*) that the highest eigenvalue distribution could be approximated by a rescaled and shifted Tracy-Widom distribution. However, little is known below that bound, and nothing is known when sparsity varies in columns. We discovered that this phenomenon can be also observed in real data; it can be detected by removing the potential biological signal by permuting the cell expression values in each gene independently. Figure 2B shows examples of the coordinate distribution of localized and delocalized vectors. Localization can be identified as deviations in the square of the components of eigenvectors from the expected distribution (Beta, that approximates to Gaussian in high dimensions, as in the single-cell data), by the Shannon entropy, or by the Inverse participation ratio (IPR) (Supplementary Figure 4, Methods). To show this behavior, we randomized a 95% sparse cell-gene expression matrix corresponding to 6,573 human PBMC cells from reference (*38*) and analyzed the statistics of its eigenvalues and eigenvectors. Although the bulk of the eigenvalue density seems to follow a MP distribution, it is easily seen that deviations on the upper edge appear that are not consistent with the expected Tracy-Widom distribution and moreover that localization of corresponding eigenvectors occurs (Figure 2C, Supplementary Figures 5 and 6). The localization phenomenon due to sparsity generates artifacts that could potentially be interpreted as true signal in standard application of PCA. For instance, outlier points can be detected in the highest components of sparse random data (Figure 2D). Another effect of sparsity is the artifactual generation of an "elbow" in randomized data sparse data (Figure 2E). These effects can be suppressed by eliminating the localized vectors, generating a more homogeneous distribution in the lower dimensional representation reflecting the random nature of the data.

**Simulations and comparison with alternative approaches**

We now proceed to evaluate the performance of the use of sparsity induced localization correction and random matrix statistics for the identification of potential relevant biological signals (Figure 3). We first perform two sets of simulations: a single-cell RNA sequencing simulation of six cell populations using Splatter (*39*) (see Methods) and seven Gaussian clusters with small mean to variance ratio in each dimension. Figures 3A and 3B show t-SNE representations of the data before and after the RMT procedure and panel C the associated MP statistics. Panels 3D, 3E and 3F represents the before, after and MP statistics of a set of seven Gaussian clusters. The first example illustrates the challenge of identifying structures based on t-SNE plots before performing the RMT procedure (Figure 3A); in contrast, after the procedure we see clearly separated clusters (Figure 3B). The test based on the MP statistics correctly identifies the six components associated with the six simulated clusters. The second Gaussian simulation correspond to a regime where the identification of clusters could be challenging due to low difference in means compared to the variance (Figure

3D). Again, the RM procedures clearly identified the cluster structure (Figure 3E, 3F).

We now perform a comparison in terms of cell-phenotype cluster resolution with some published algorithms. For that purpose, we are using the data sets (*38*) (human PBMC) and (*40*) (mouse cortex) described in the previous section. As explained in the previous section, these references together with the analysis done in (*11*) about (*40*) have cells already labeled in terms of phenotype. We claimed in previous section that our RMT method is able to clean system noise such that the cell-phenotype clusters are better resolved. This noise is partially generated due to the missing values in single-cell experiments. For that reason, we compare with the two main approaches in the field that address this question: imputation (MAGIC (*8*) and scImpute (*10*)) and zero-inflated dimensionality reduction (ZIFA (*13*) and ZIMB-WaVE (*9*)). For completeness, we also have compared with the raw data, with a selection of genes based on higher variance (top 300 genes) and with Seurat (*41*). The comparison is performed using the knowledge of cell phenotypes in refs. (*11, 38, 40*) and by computing the mean silhouette score in the reduced space; higher values would indicate a better (less noisy) cell-phenotype cluster resolution. In Figure 3G-3J we represent the mean silhouette score as a function of the reduced space number of dimensions (number of PCs) for 13 PBMC cell-phenotypes described in (*11*)(Figure 3G) and for 7 (Figure 3H), 15 (Figure 3I) and 26 (Figure 3J) marked mouse cortex cell populations described in (*40*). We have selected the 1,500 most signal-like genes using RMT and we can observe how RMT outperforms other methods in the identification of known marked populations. Notice also how this becomes more dramatic as we increase the number of populations (Figures 3H to 3J, Supplementary Figures 7-9). Although this exercise is done with known populations in order to give a comparative quantitative measure, from Figures 3G-3J we can also conclude that RMT method is a suitable one to better disentangle cell populations by noise removal and hence to find new potential cell populations. Moreover, it can also be noticed that the RMT method is increasingly better for higher number of PCs. This last feature is particularly important since in the future, due to the improvements in resolution and number of cells, the number of required dimensions (PCs) for an accurate analysis will grow.

**Application to diverse single-cell transcriptomic data sets**

In this section we present in detail the RMT analysis of the two marked single-cell data: 6,573 human PBMC cells from reference (*38*) (Figure 4, Supplementary figure 10) and 3,005 mouse cortex cells from reference (*40*) (Figure 5, Supplementary figure 11). Both data sets have the ground truth labels for each cell that explaining the different cell populations existing in both systems. In the case of reference (*38*) we have additionally used the results obtained in (*11*) where the authors analyze data in (*38*) and produce a larger number of labels (corresponding to more cell identities). We ran the RMT algorithm that first eliminates the genes that introduce artifacts due to sparsity, then determines how many PCs are pure noise based on the prediction from the universal MP distribution (Figures 4A and 5A, see Methods). Finally, by projecting in signal and in different regions of noise (see Methods) the algorithm determines how much each gene is responsible for signal or noise based on a chi-squared test for the variance (Figures 4B and 5B, see Methods). The panel C in Figures 4 and 5 corresponds to the mean silhouette score as a function of the chi-squared test of variance. This is related with panel B where the chi-squared test for variance determines the number of genes mostly responsible for signal. The idea expressed in panel C is that, due to the fact that we already know the identity of cells in each data set and provided that cells cluster by (phenotypic) similitude, higher values of mean silhouette score implies a better (and hence cleaner) cell cluster resolution. Panels B and C together demonstrate the utility of the RMT approach: by selecting the genes that carry the biological signal we eliminate the noise of the system. In panels D and E of Figures 3 and 4 we have performed a hierarchical clustering of the clean system and projected the clusters into a t-SNE plot in order to better visualize it. We also performed a comparison with the cell identities defined in references (*11, 38, 40, 42*) (D panels), confirming that we can recover the populations in these references and adding some more potential subpopulations (E panels). In particular, for the PBMC case we have found two subpopulations for dendritic cells that were not previously identified in the original work (*38*) and subsequent further refinements (*11*), that expressed markers identified in an independent study (*43*) (Supplementary Figure 16). In addition, we identified three for CD14 monocytic cells, two for B-activated cells and two for

T-activated cells (see Methods for the discussion of the differential expression analysis).

**Discussion**

In this manuscript, we demonstrate the effectiveness of tests based on (sparse) Random Matrix Theory for studying the spectrum of the covariance matrix of single-cell genomic data. We show that while most of the spectrum follows the expectations from RMT (95%), deviations become apparent both in the distribution of density of eigenvalues and the maximum eigenvalue and in eigenvector localization. We further show that most of these effects are artifacts due to the sparsity of the data including sparsity induced localization; this accounts for 3% in average of the remaining eigenvalues. The final 2% of the eigenvalues could then be attributed to true biological signal. Eigenvector localization after eliminating the sparsity effects points to groups of cells where information tightly concentrates, indicating common transcriptional programs. Sparse Random Matrix Theory and associated eigenvector localization correction provides a powerful tool to identify this signal and produce a low rank representation of single-cell data that may be used for further interpretation. Additionally, we should point out that the universality we observed in Wishart/covariance matrices is also observable in the spectra of graph Laplacians (including sparse graphs (*44*)) and kernel random matrices (*45*), which are used in other single-cell analytic techniques, suggesting that the approach followed here could be applied more broadly.

Code for the algorithm and denoising pipeline is publicly available on https://rabadan.c2b2.columbia.edu/html/randomly/ .

## ACKNOWLEDGEMENTS

This work was funded in part by R01CA185486-01, R01 CA179044-01A1, U54 U54CA209997, NIH U54 CA193313 and a Chan-Zuckerberg pilot grant. We want to thank Francesco Brundu, Tim Chu, Oliver Elliott, Ioan Filip, Antonio Iavarone, Zhaoqi Liu, and Richard Wolff. We would like to especially thank Ivan Corwin for helpful discussions about mathematical aspects and implications of this work.

# FIGURE LEGENDS

**Figure 1**. **Quasi-universality of single-cell sequencing data**. A) The Wigner surmise distribution captures the spacing between eigenvalues of Wishart matrix across single-cell RNASeq experiments. B) Departures from random matrix universal distributions indicate interesting potential biological signals. In red is the non-parametric Marchenko-Pastur distribution. Deviations from universality can be found by analyzing the larger eigenvalues in relations to the expected Tracy-Widom distribution. C) Departures from universality amount to near 5% of eigenvalues. However, most of these can be explained by the sparsity of data, suggesting that Sparse Random Matric Theory can provide a better model to understand single-cell sequencing data. Truly potential biological signal amounts to only ~2% of eigenvalues.

**Figure 2**. **Sparse Random Matrices and Eigenvector Localization can model single-cell data.** A) Deviations from Tracy Widom distributions can be easily appreciated in sparse matrices in the mean and the median of the highest eigenvalue at a function of fraction of non-zero values (p). In this case, the 100 by 100 random matrices are drawn a mixture of a normal and a Dirac-delta at zero. Similar results are obtained with other sparse distributions (Poisson, Supplementary Figure 3), B) The deviations from Tracy-Widom have been associated to the phenomenon of eigenvector localization: while in non-sparse matrices eigenvector are randomly distributed in a sphere (Haar measure in orthogonal groups), in sparse matrices eigenvectors are localized along some directions. C) Localization can be identified as deviations in square of the components of eigenvectors from the expected distribution that approximates Gaussian in high dimensions. D) The localization phenomenon due to sparsity generates artifacts as outliers that can bias the lower dimensional representations. Eliminating the localized vectors generates a more homogeneous distribution in the lower dimensional representation reflecting the random nature of the data. E) The effects of sparsity can also be appreciated in the classical elbow plots: sparsity can introduce an artifactual elbow in randomized data.

**Figure 3**. **Simulations and comparison with alternative approaches for single-cell denoising.** A) t-SNE representation of a six cell populations single-cell simulation using Splatter (*39*), **B)** results after processing through the RMT procedure, **C)** Marchenko-Pastur prediction and identification of the relevant components. **D)** t-SNE representation of a single-cell simulated of seven Gaussian cell populations, **E)** results after processing through the RMT procedure, **F)** Marchenko-Pastur prediction and identification of the relevant components. **H)** Mean Silhouette score for different methods as a function of the reduced space number of dimensions (number of PCs) for the case of 13 PBMC cell-phenotypes described in (*11*). **I)** Mean Silhouette score for different methods as a function of the reduced space number of dimensions (number of PCs) for the case of 7, 15 **J)** and 26 **K)** mouse cortex cell-phenotypes described in (*40*).

**Figure 4**. **Application to PBMC single-cell expression. A)** Marchenko-Pastur prediction and identification of relevant components. **B)** Study of the chi-squared test for the variance (normalized sample variance) in signal and noise gene projections. In the left subpanel the distributions correspond to a projection of genes into the 83 signal eigenvectors (corresponding to the 83 eigenvalues of panel (a)) and the projection into the 83 lowest and 83 largest Marchenko-Pastur eigenvectors. There is also a projection into 83 random vectors. Finally, the lines show how gamma-functions can fit the distributions discussed. The right subpanel shows the number of relevant genes in terms of the test above discussed together with a false discovery rate. The higher is the chi-squared test for variance the less genes are responsible for signal. **C)** Evolution of the mean Silhouette score as a function of the statistical test discussed in panel (b). The score grows at the beginning because genes responsible for noise are eliminated as the sample variance increases. At a certain threshold the score starts dropping because the number of genes is too small to define the cell clusters. Nullified case corresponds to genes projected into signal before selecting the signal-like genes (the RMT uses this case). **D)** Hierarchical clustering of the PBMC cells after using RMT method to eliminate noise and comparison with the cell populations found in (*38*) and (*11*). The number of genes signal-like selected is 1500. **E)** Visual representation of the previous panel through a t-SNE plot.

**Figure 5**. **Application to mouse cortex single-cell expression. A)** Marchenko-Pastur prediction and identification of relevant components. **B)** Study of the chi-squared test for the variance (normalized sample variance) in signal and noise gene projections. In the left subpanel the distributions correspond to a projection of genes into the 103 signal eigenvectors (corresponding to the 103 eigenvalues of panel (a)) and the projection into the 103 lowest and 103 largest Marchenko-Pastur eigenvectors. There is also a projection into 103 random vectors. Finally, the lines show how gamma-functions can fit the distributions discussed. The right subpanel shows the number of relevant genes in terms of the test above discussed together with a false discovery rate. The higher is the chi-squared test for variance the less genes are responsible for signal. **C)** Evolution of the mean Silhouette score as a function of the statistical test discussed in panel (b). The score grows at the beginning because genes responsible for noise are eliminated as the sample variance increases. At a certain threshold the score starts dropping because the number of genes is too small to define the cell clusters. Nullified case corresponds to genes projected into signal before selecting the signal-like genes (the RMT uses this case). **D**) Hierarchical clustering of the PBMC cells after using RMT method to eliminate noise and comparison with the cell populations found in (*38*) and (*11*). The number of genes signal-like selected is 1500. **E)** Visual representation of the previous panel through a t-SNE plot.

Figure 1

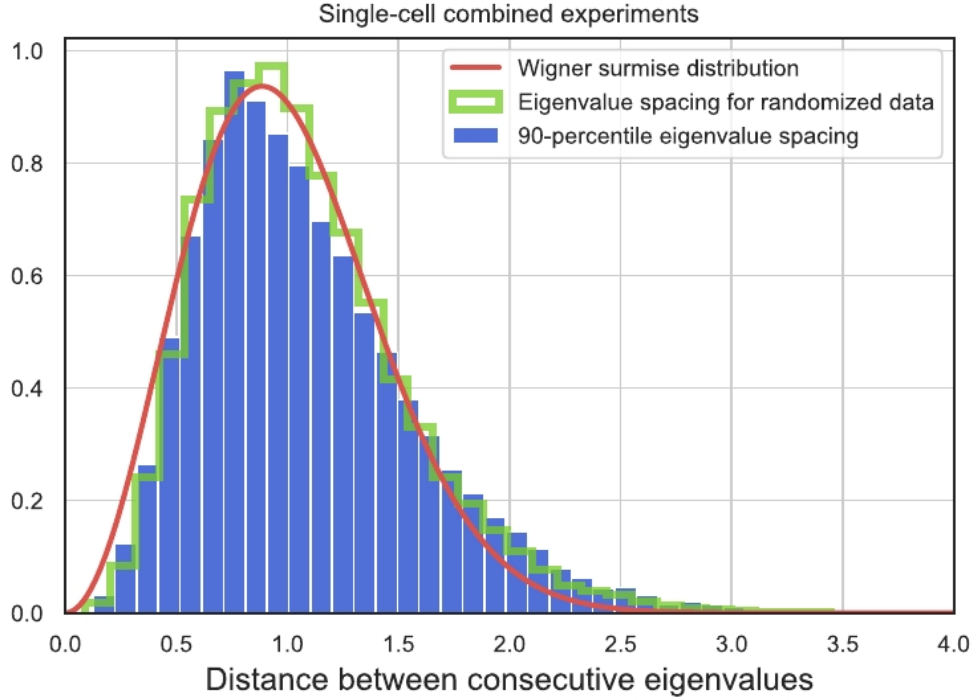
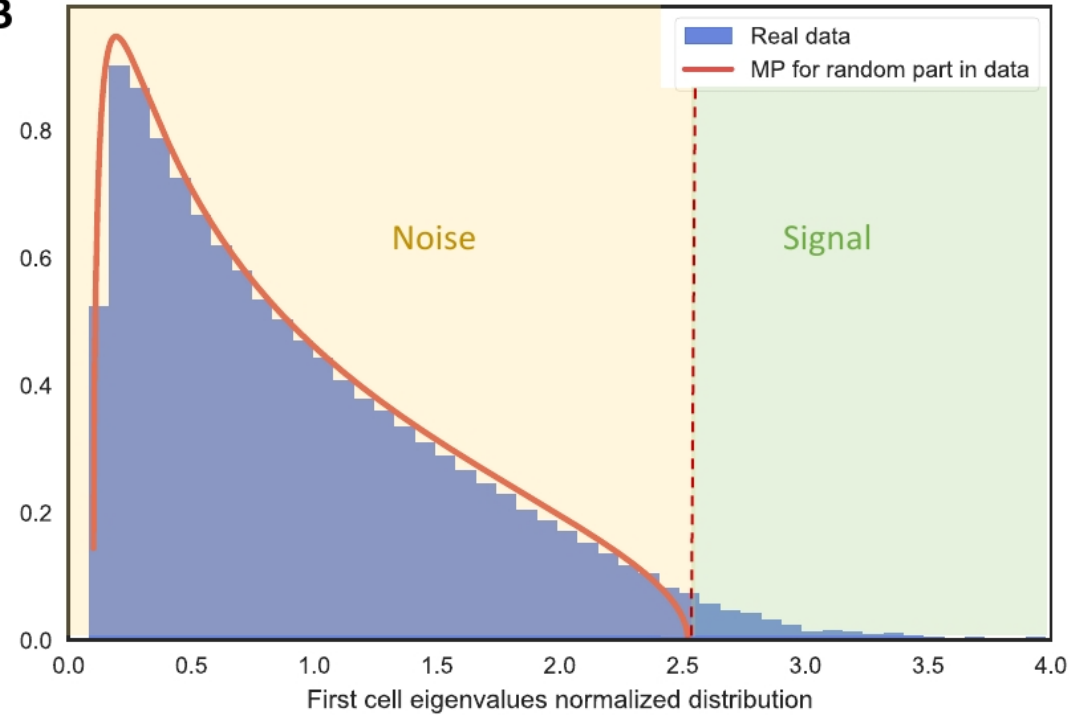
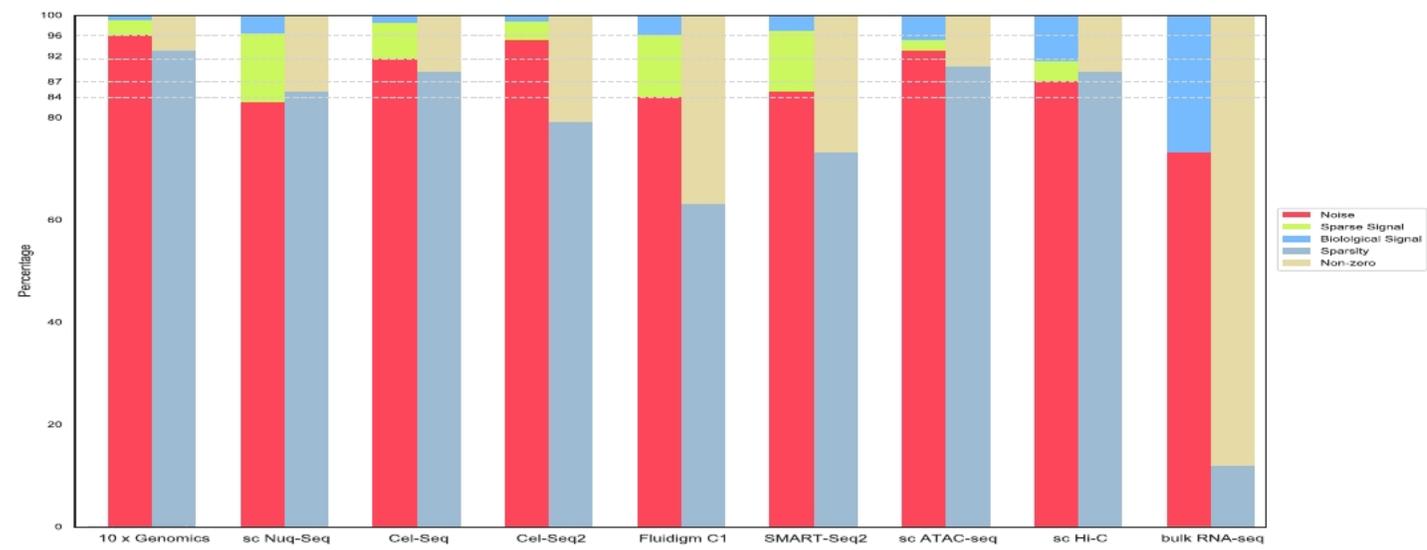

Figure 2

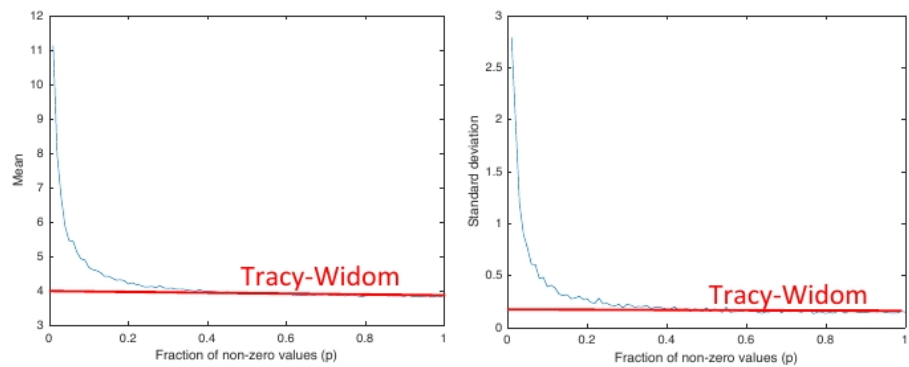
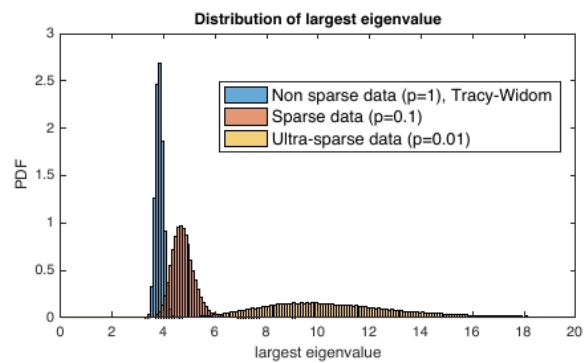
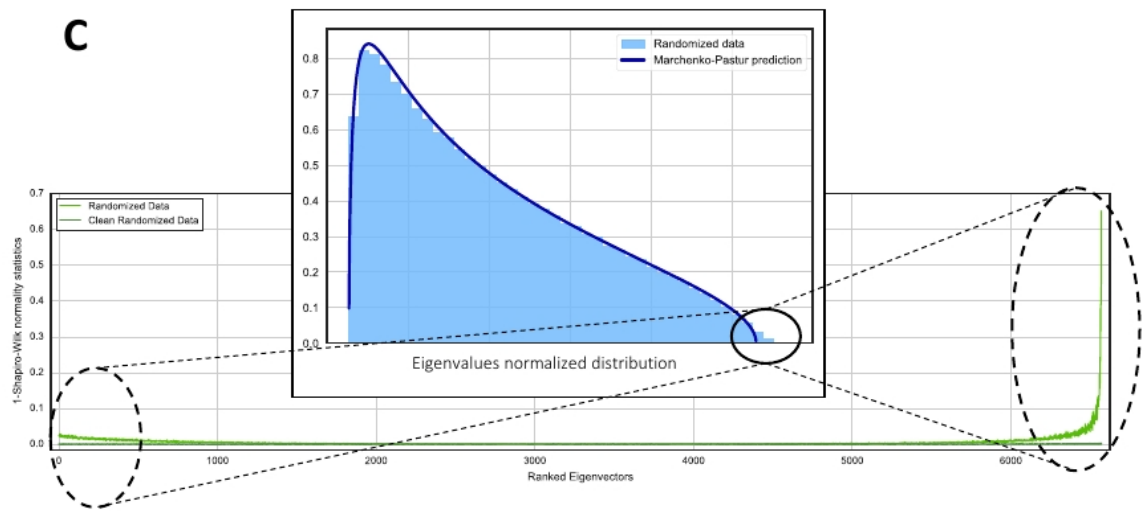
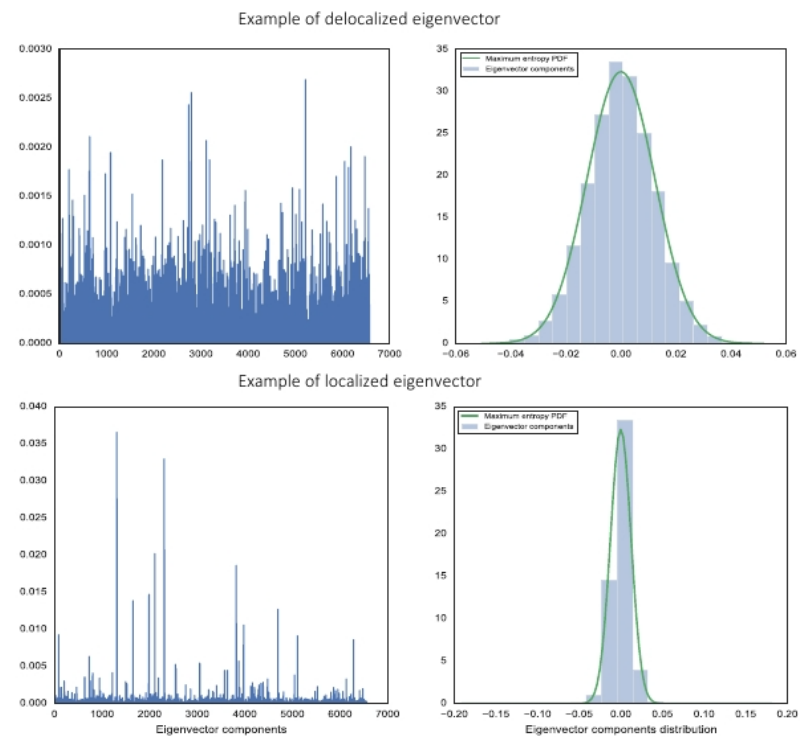
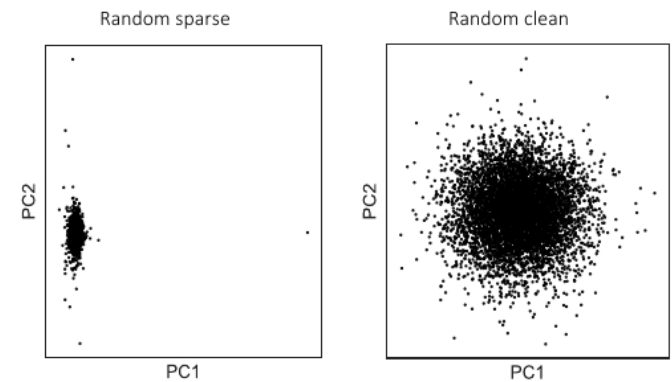
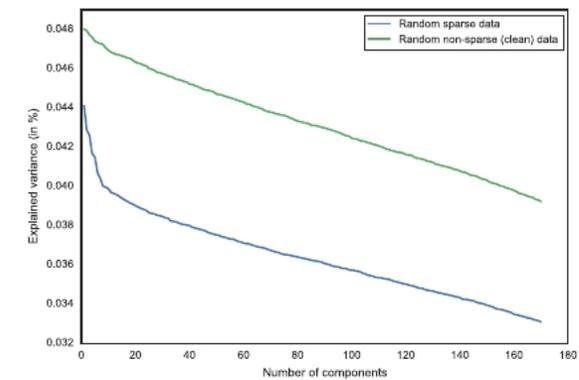

Figure 3

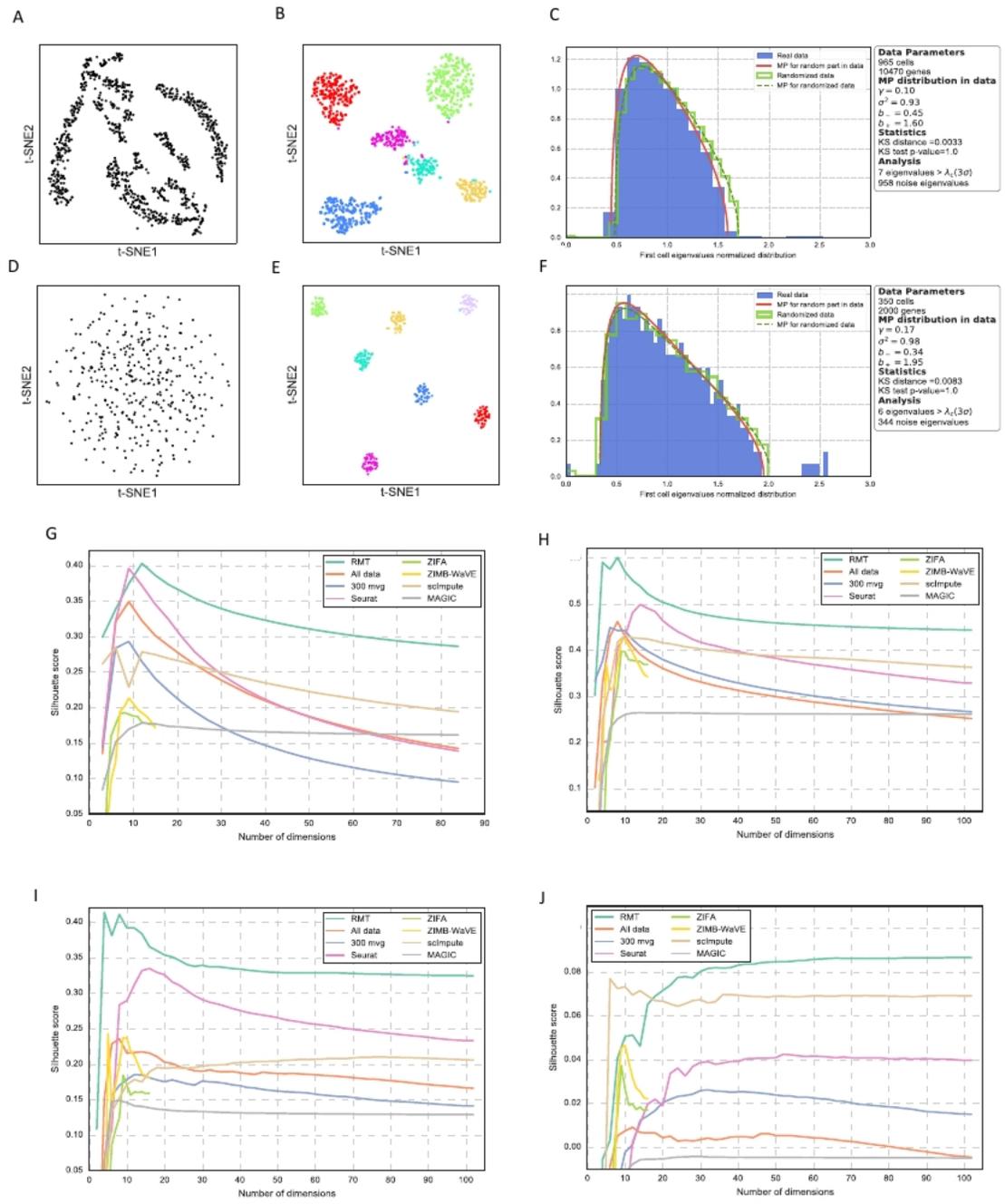

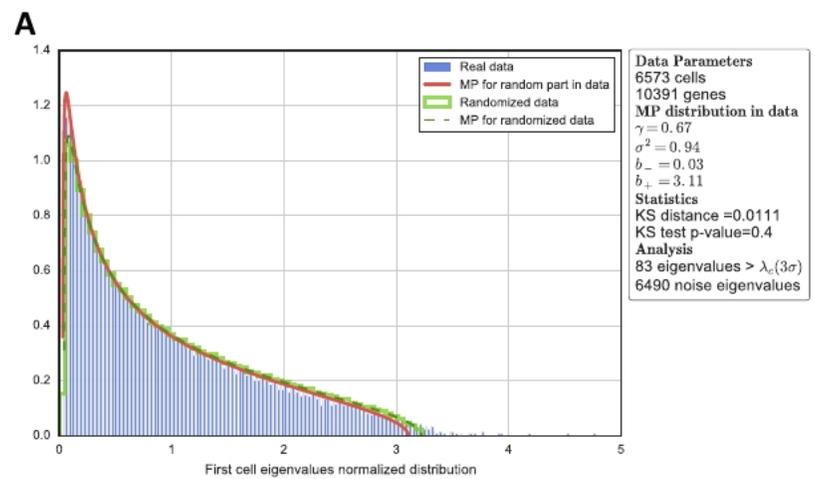
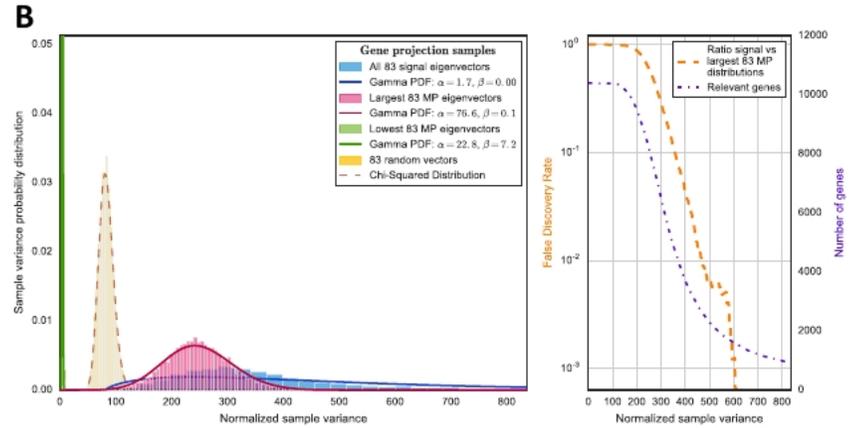
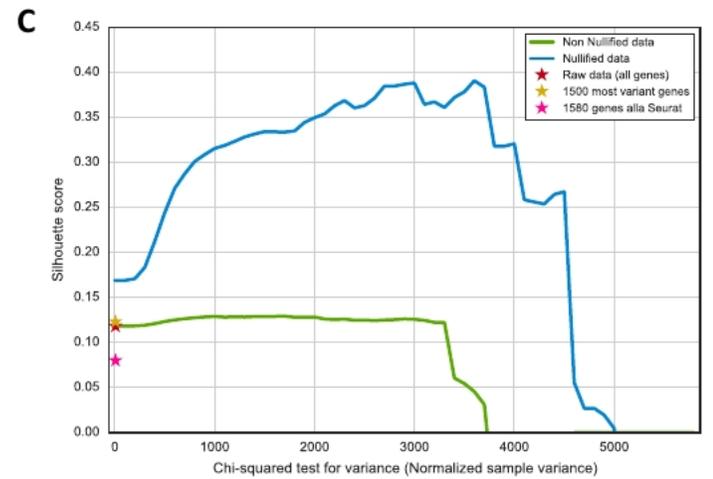
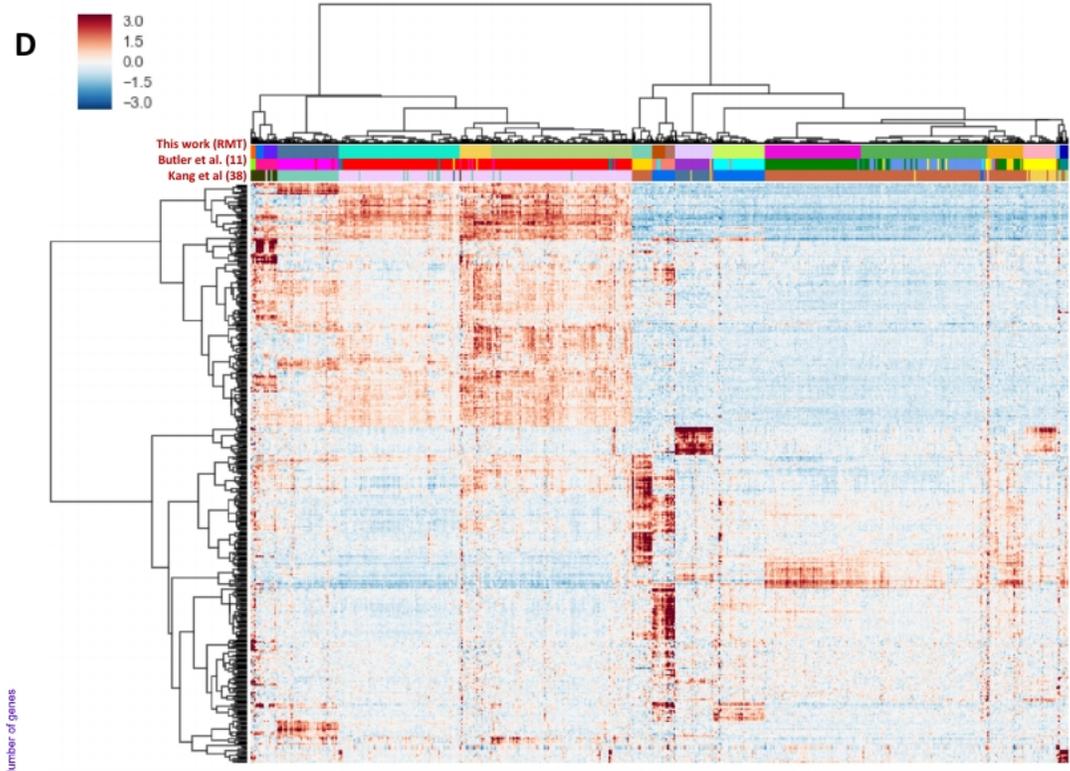
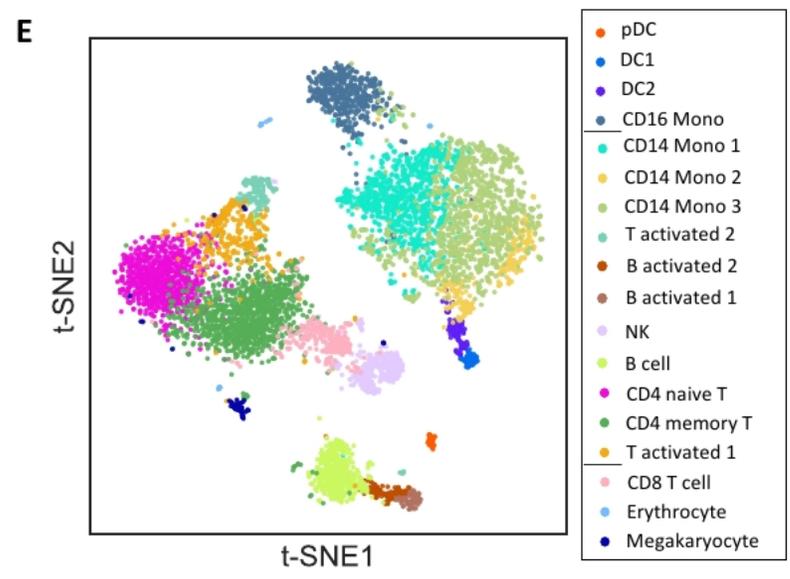

Figure 5

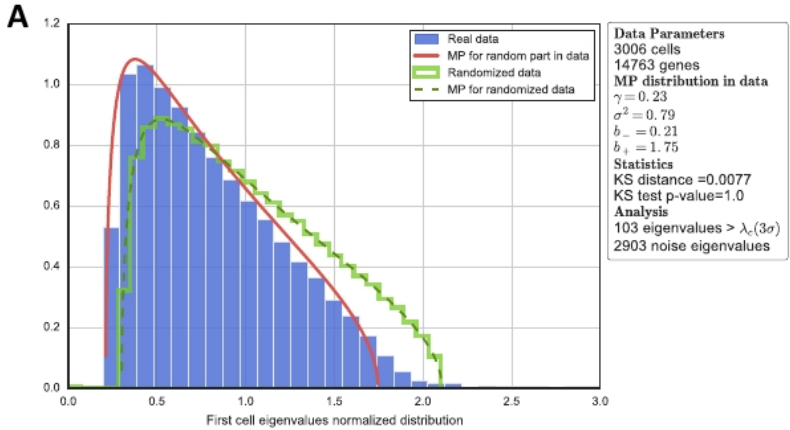
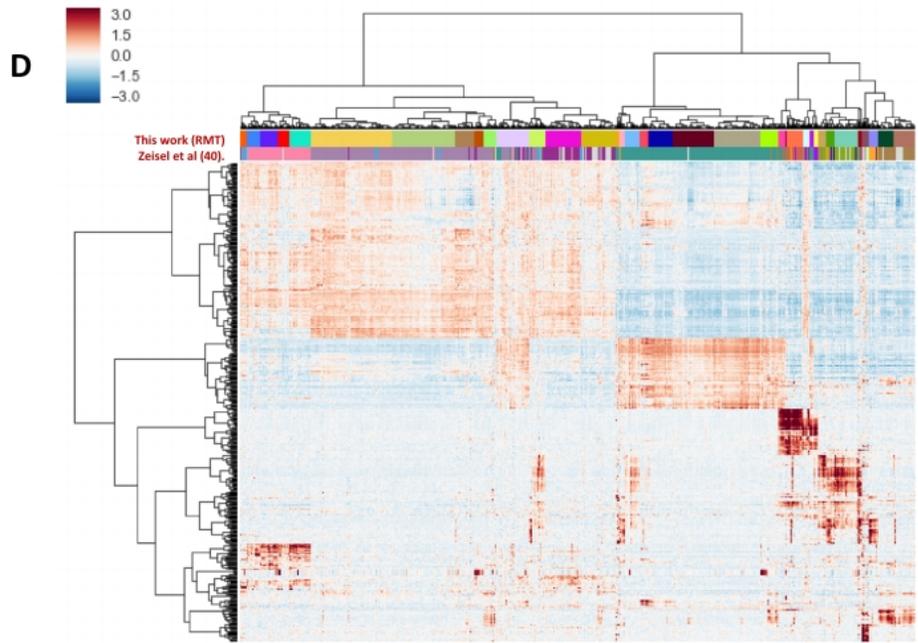
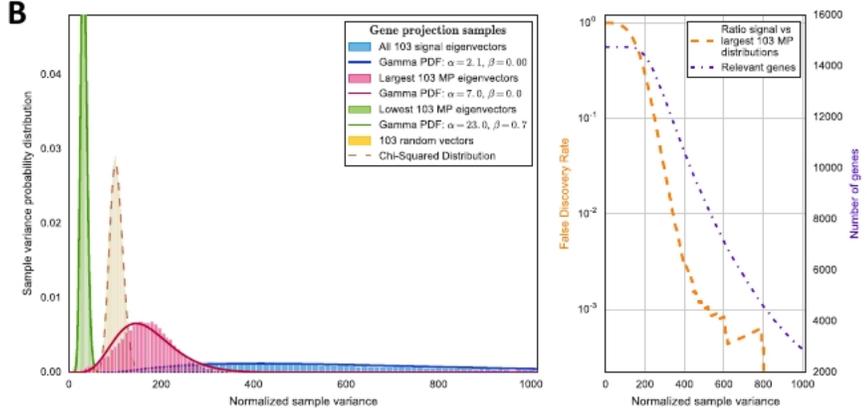
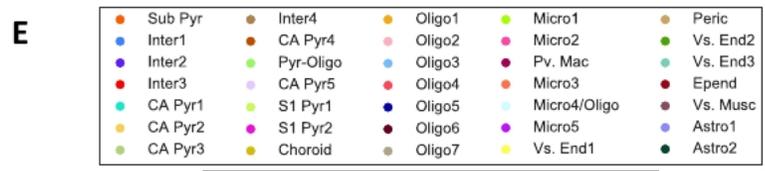
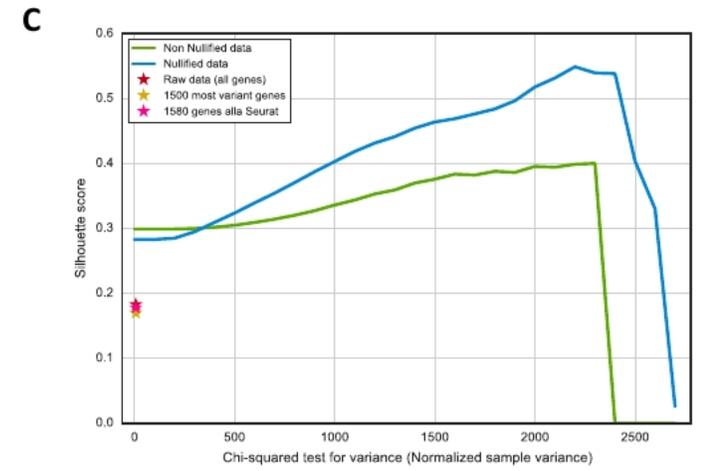
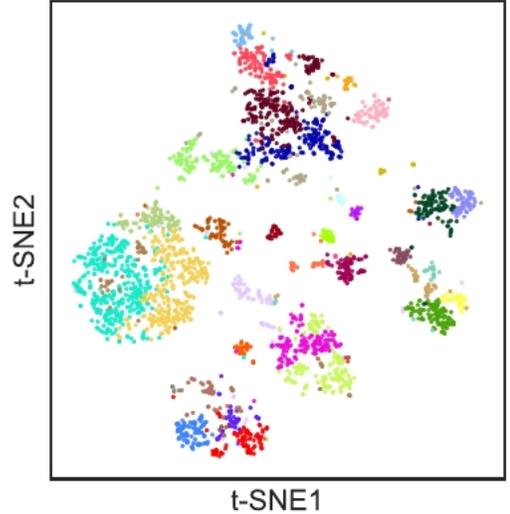

## METHODS

**Introduction to eigenvalue statistics in covariance random matrices**

Given a $N \times P$ matrix $X$, each column is independently drawn from a distribution with mean zero and variance $\sigma$, the corresponding Wishart matrix is defined as

$$W = \frac{1}{P} XX^T$$

The eigenvalues $\lambda_i$ and normalized eigenvectors $\psi_i$ of the Wishart matrix where $i = 1, 2, \ldots N$ are given by the following relation:

$$W\psi_i = \lambda_i \psi_i$$

If $X$ happens to be a random matrix (a matrix whose entries $x_{ij}$ are randomly sampled from a given distribution) then $W$ becomes a random covariance matrix and the properties of its eigenvalues and eigenvectors are described by Random Matrix Theory (RMT). In the case of the random distribution being normal with mean 0 and variance 1, one can refer to this as a Wishart ensemble. One of the most interesting properties RMT is the so-called *universality* of the eigenvalue local and global statistics. The global statistics consist of the study of eigenvalue distribution of $\mathcal{O}(N)$ number of eigenvalues. On the contrary, local statistics study the behavior of a small number of eigenvalues, like the distribution of distances between neighboring ordered eigenvalues, the distribution of the largest and smallest eigenvalues and the correlation functions. A property is called universal if it only depends on the symmetry properties that define the ensemble and not on specific details of the underlying probability distribution beyond the first few moments. Universality properties arise both at the local and global scales in the limit $N \to \infty$, $P \to \infty$, $\gamma = \frac{N}{P}$ fixed.

The global statistics are determined by the calculation of the eigenvalue density or empirical density of states:

$$\rho(\lambda) = \langle \frac{1}{N} \sum_{i=1}^{N} \delta(\lambda - \lambda_i) \rangle$$

which for the Wishart matrices converges in the limit $N \to \infty$, $P \to \infty$, $\gamma = \frac{N}{P} \leq 1$ to the so-called Marchenko-Pastur (MP) distribution:

$$\rho_{MP}(\lambda) = \frac{1}{2\pi\gamma\sigma^2} \frac{\sqrt{(a_+ - \lambda)(\lambda - a_-)}}{\lambda} \mathbb{I}_{[a_-, a_+]}$$

where

$$a_\pm = \sigma^2 (1 \pm \sqrt{\gamma})^2$$

If $N \to \infty$, $P \to \infty$, $\gamma = \frac{N}{P} > 1$, then the Marchenko-Pastur (MP) distribution has a delta function centered at zero:

$$\rho_{MP}(\lambda) = \frac{1}{2\pi\gamma\sigma^2} \frac{\sqrt{(a_+ - \lambda)(\lambda - a_-)}}{\lambda} \mathbb{I}_{[a_-, a_+]} + (1 - \frac{1}{\gamma})\delta(0)$$

The parameter $\sigma$ represents the variance of the probability distribution that generates each element in the random matrix ensemble. In Supplementary Figure 14 there is a graphical representation of the eigenvalue density with two regimes for the ordered eigenvalues: the bulk and edges (largest and smallest eigenvalues). The emergence of MP density is already a form of universality, because the density of eigenvalues is asymptotically the same regardless of the details of the probability distribution of the individual matrix elements. If the first two moments are fixed to be 0 and 1 and the distribution has a sufficient number of finite moments, the universality property of the local statistics of the covariance matrix is satisfied without requiring that the entries of the underlying random matrix be i.i.d. (*1, 2*).

The local statistics of eigenvalues are based on Wigner's original observation concerning the distribution of the distances (gaps) between consecutive eigenvalues, colloquially known as the Wigner Surmise:

$$P(s) \approx \frac{\pi s}{2} e^{\left(-\frac{\pi}{4}s^2\right)} ds$$

where $s = (\lambda_j - \lambda_{j-1})/D$ and $D$ is the mean spacing among eigenvalues.

Local universality has been shown in two flavors (*1-4*):

- Universality of local eigenvalue statistics:
    1. Bulk universality is the celebrated Wigner-Dyson-Gaudin-Mehta conjecture. It says that regardless of the probability distribution, the $n$-point correlation functions are described by the sine kernel:

    $$K\left(\frac{\lambda_i}{N}, \frac{\lambda_j}{N}\right) = \frac{\sin \pi(\lambda_i - \lambda_j)}{\pi(\lambda_i - \lambda_j)}$$

    Notice that the kernel does not factorize and therefore shows the strong correlation among eigenvalues. On the other hand, given that the distribution of gaps can be computed from the correlation

functions (*5*), bulk universality also explains the Wigner Surmise-like universality.

2. Largest eigenvalue universality: the behavior of the largest eigenvalue $a_+$ is such that

$$Prob(\lambda_{max} < t) \approx F_\beta(N^{2/3}(t - a_+))$$

where $F_\beta$ in the Wishart case is a function known as the Tracy-Widom distribution. A similar argument can be made for the smallest eigenvalue $a_-$.

- Eigenvector delocalization: this property implies that the norm of the eigenvectors $\psi_i$ is equally distributed among all their components $\alpha$:

$$\left|\psi_i^{(\alpha)}\right| \sim \frac{1}{\sqrt{N}}$$

**Introduction to eigenvector localization**

Let us now consider the case of perturbed random matrices, i.e. matrices that contain a random part plus a perturbation that partially breaks the randomness in some direction. The spike model of Johnstone provides a simple example where a finite rank perturbation is added to a large random matrix (*6*). If the perturbation is below a certain critical value the largest eigenvalue follows the Tracy-Widom distribution of the unperturbed matrix. However, if the perturbation is larger than the critical value, the largest eigenvalue may separate from the bulk MP distribution and present Gaussian fluctuations (*7*). This is the so-called BBP (from Baik-Ben Arous-Peche) phase transition. Subsequent works, like (*8*), have shown that there is no universality in the fluctuation pattern for eigenvalues that separate from the bulk—i.e. the parameters of the distribution of the fluctuation depends on the perturbation features. A similar phase transition was found at the eigenvector level by (9). Eigenvectors associated with the eigenvalues out of the bulk get localized: the norm gets concentrated in a small number of coordinates containing information about the original perturbation. In Figure 2B (left panels), we show an example of localized and delocalized eigenvectors. The x-axis represents the order of the components, and the y-axis the squares of the components' values $\left(\psi_i^{(\alpha)}\right)^2$. Notice that the eigenvectors are normalized, such that

$$\sum_{\alpha=1}^{N}\left(\psi_i^{(\alpha)}\right)^2 = 1$$

The delocalized-localized phase transition is an example of the famous Anderson localization phase transition that was first observed in quantum disordered systems (10). It has been observed that, in the delocalized phase, the eigenvalue statistics are governed by RMT.

Interestingly, the distribution of components for delocalized eigenvectors can be easily estimated by considering them as vectors on a unit sphere of dimension N-1. The distribution of their components is then given by

$$f(\psi) = (1-\psi^2)^{\frac{N-3}{2}}$$

that in case of large N approximates a Gaussian distribution with mean zero and 1/N variance (see Figure 2B)

$$f(\psi) \sim \frac{N}{\sqrt{2\pi}} e^{\left(\frac{-N\psi^2}{2}\right)}$$

This can be made more precise in terms of information theory, by noticing that delocalized eigenvectors do not carry any information, while localized vectors are correlated with the original perturbation. This insight proves very useful when attempting to distinguish between noise and signal in any complex system, biological systems in particular. The noise in the data will correspond to the part of the spectrum that can be described in terms of RMT.

In Supplementary Figures 5A and 6A we implement a Shapiro-Wilk normality test on the coordinates of the eigenvector from two test datasets, allowing us to separate signal from noise. In addition, as a mean to confirm the results, we perform two alternative statistical tests. The first test is based on information theory and its results are shown in Supplementary Figures 5B and 6B. We have compared the Shannon entropy for each eigenvector with randomized data. The second test is based on applications to financial data (*11*) and the result can be seen in Supplementary Figures 5C and 6C. In this case we are calculating the inverse of the Inverse Participation Ratio (IPR):

$$IPR_i = \sum_{\alpha=1}^{N} \left(\psi_i^{(\alpha)}\right)^4$$

The inverse of the IPR quantifies the number of eigenvector components that contribute significantly. The results are equivalent to those obtained with the other statistical tests.

**Sparse Random Matrix ensembles and sparsity induced localization**

Sparse Random Matrix Theory (sRMT) ensembles are a class of random matrices with a fraction of non-zero elements, p. In contrast to the RMT's presented in the first section of the Methods, sRMT's exhibit localized eigenvectors, a phenomenon that we call here sparsity induced localization.

The universality properties of covariance sRMT have been studied in (*12*), (*13*) and (*14*). The local statistics of eigenvalues preserve the bulk and largest eigenvalue universalities. The main difference with non-sparse RMT is the global statistics (*15*) and the presence of localized eigenvectors (*16, 17*). Regarding the global statistics, there could be significant deviations from the original MP and Tracy-Widom distributions, depending on the fraction of non-zero values p (Figure 2A, 2C). In these figures we are using sparse ensembles from a mixture of Gaussian distribution, Dirac delta distribution centered at zero and Poisson distribution applied to the randomized dataset (*18*).

The presence of localized eigenvectors is a very important feature. In the bottom panel of Figure 2C, the correlation between the MP deviations and the presence of localized eigenvectors is shown, by using the gaussianity test discussed above. In Supplementary Figures 4A and 4B, we evaluated the localization of eigenvectors using two other tests: Shannon entropy and IPR, previously described. The three tests show sparsity induced localization. This sparsity induced localization is not associated with any biologically relevant information. Sparsity induced localization can introduce artifacts as outliers in PCA and artifactual elbow plots (Figure 2D, 2E and Supplementary 4C). We have also performed a comparison between the sparse dataset and the one after removing sparsity in Figures 2C, 2D and 2E applied to the randomized dataset (*18*). Colors distinguish between sparse and clean data. The same comparison has been performed for the original datasets (*18*) (Supplementary Figure 5) and (*19*) (Supplementary Figure 6).

**Algorithm description for denoising of single-cell data**

We outline three major steps in the denoising of single-cell data algorithm on the example of PBMC dataset by Kang et al. (*18*), and illustrate in the supplementary Figure 13.

- **Preprocessing**

The goal of preprocessing is to remove genes that create artifacts, due to the sparse nature of the data. Gene expression values for each cell were divided by the total number of transcripts and multiplied by $10^6$. These values were then log2 transformed. After, the single-cell data matrix X is Z-score normalized, such that every gene has mean 0 and standard deviation 1. A randomized matrix is obtained via random permutation of cells for every gene independently, to destroy potential correlations. We project the expression of each gene onto the eigenvector basis of the randomized matrix. To assess normality, we evaluated several related methods: Kolmogorov-Smirnov, Anderson-Darling, and the Shapiro-Wilk test, all providing similar results. In this manuscript, we use the Shapiro-Wilk statistics comparing to genes that express less than certain number of transcripts (in this manuscript, 7 transcripts by default) (see Supplementary Figure 13(a)). Genes that have Shapiro-Wilk statistic higher than the minimum statistic of the sparse genes with less that 7 transcripts are considered to be abnormal and are removed from the further analysis. Alternatively, as the p-value is a monotonic function of the Shapiro-Wilk statistic, one can impose an equivalent cut-off on p-value, correcting for multiple hypotheses.

- **Marchenko-Pastur parameter estimation**

After the identification and removal of abnormal genes, Wishart matrix of the preprocessed data is constructed, and a full set of eigenvalues and eigenvectors is computed using standard Singular Value Decomposition (SVD) algorithms. Full set of eigenvalues of the Wishart matrix are required to estimate the parameters of the MP distribution. Gradient descent iterative search is implemented to find an optimal fit of the MP distribution with eigenvalues of the randomized matrix as an initial step in the iterative process. Eigenvalues that fit MP distribution are considered to be consistent with the noise (see Supplementary Figure 13(b)). Eigenvalues above Tracy-Widom critical eigenvalue are considered to be associated with biological signal.

- **Gene Selection**

To select genes that are the most consistent with biological signal, we analyze the variance of every gene projected onto signal eigenvectors and compare it to the largest variance that can be attributed to noise. We project genes onto four subsets of equal size of the eigenvectors of the Wishart matrix in question: signal eigenvectors that correspond to the eigenvalues above Tracy-Widom critical eigenvalue; eigenvectors right below the critical eigenvalue of Tracy-Widom distribution; eigenvectors of lower spectrum of MP distribution; and equal number of eigenvectors in the middle of MP distribution spectrum.

Our goal is to infer the maximum and minimum variance that genes can have due to noise. We select the most variant genes across signal eigenvectors versus noise eigenvectors. Note that these genes are different from the most variant genes across all the eigenvectors in general.

Eigenvectors in the middle of MP distribution spectrum, are considered to be the most compatible with noise. Variance distibution of the genes projected onto these eigenvectors can be modeled using standard $\chi^2$ distribution. Variance distribution of genes projected onto the eigenvectors corresponding to the set of largest eigenvalues of the MP distribution has standard deviation larger than that of $\chi^2$ distribution. Variance distribution of genes projected onto the eigenvectors corresponding to the set of lowest eigenvalues of the MP distribution has standard deviation smaller than that of $\chi^2$ distribution. These variance distributions can be modelled using Gamma distributions. We estimate the parameters of the Gamma distributions and $\chi^2$ distribution using standard maximum likelihood estimation procedure. To select genes, we compare variance of genes across signal eigenvectors and right spectrum of MP distribution (see Supplementary Figure 13(c)). We establish False

Discovery Rate FDR (0.001 by default). Genes that have a ratio of variance across subset of MP eigenvectors right below the Tracy-Widom cut-off (largest variance associated with noise) and across subset of signal eigenvectors below the FDR are selected.

As a result of the denoising algorithm, the eigenvalues compatible with noise are nullified and genes that have variance across signal eigenvectors compatible with noise are removed (controlled by free parameter FDR).

**Datasets**

In Figures 2C, 4 we are using Kang et al. (*18*) GSE96583 with their labels and a second set of labels (those referred as control by the authors) from Butler et al. (*20*). For Figure 5, we used Zeisel et al. dataset GSE60361 and annotation from (*19*).

For Figure 1C, we are using the following datasets:
- Count matrix for the SMART-Seq2 data set (*21*) was obtained under the GEO accession number GSE81682.
- Count matrices for CelSeq and CelSeq2 data sets were obtained from accession numbers GSE81076 and GSE86469 correspondingly.
- Count matrix for Fluidigm C1 technology was obtained from accession number GSE86469.
- Hi-C data was obtained from accession number GSE84290.
- Data for the ATAC-seq data was obtained under the accession number

GSE65360.

- Raw Nuq-seq data was obtained from the Gene Expression Omnibus with accession number GSE84371.

- Data for bulk RNAseq GBM was obtained from TCGA firehose portal (illuminahiseq_rnaseqv2-RSEM_genes(MD5), http://firebrowse.org/?cohort=GBM&download_dialog=true).

- For the 10x platform and human PBMC dataset, the data was obtained from 10x genomics (https://support.10xgenomics.com/single-cell-gene-expression/datasets/2.1.0/t_4k).

For Supplementary Figure 1, we are using the following datasets:

- Murine embryonic stem cell (mESC) differentiation was obtained from the NCBI Gene Expression Omnibus (GEO) database, with accession number GSE94883.

- High grade glioma dataset was taken from GSE103224.

- Kang et al. (*18*).

- Pancreas islet single-cells dataset GSE84133.

For Supplementary Figure 13, we are using Kang et al. dataset.

**Wishart matrix statistics simulations**

Single-cell RNAseq simulated dataset (Figure 3A) was generated using the Splatter R package. A mean expression level for each gene is simulated using a gamma distribution. The negative binomial distribution is used to generate a count for each cell based on these means, with a fixed dispersion parameter.

The simulation was done for 985 cells with 14472 genes for Library size (Location, Scale, Norm) 6 groups of cells with the following proportions (0.1, 0.2, 0.3, 0.2, 0.1, 0.1). The following Splatter parameters were used: Mean (Rate, Shape): (0.79, 9.58): (10, 0.69, False). Exprs outliers (Probability, Location, Scale): (0.02, 4.62, 0.91), Diff expr(Probability, Down Prob, Location, Scale): (0.1, 0.5, 0.1, 0.4), BCV (Common DISP, DOF): (0.19, 38.8), Dropout (Midpoint, shape) : (-0.085, -1.14), Paths (From, Length, Skew, Non-linear, Sigma Factor) : (0, 100, 0.5, 0.1, 0.8).

The simulation dataset in figure 3D was done by drawing random gene expression values from a multivariate normal distribution for 7 overlapping clusters with 50 cells per cluster, 100 signal genes and 2000 genes in total. Constant mean expressions for signal genes across 7 clusters were set to be [0.7, 1.7 …, 6.7].

**Comparison to other techniques**

When comparing with other methods, we are normalizing the data using $log_2(1 + TPM)$ for ZIFA (*22*), and scImpute (*23*). ZIMB-WaVE (*24*) does not need any normalization. For Seurat (*25*), we are using normalization.method = "LogNormalize" with scale.factor = 10000 and finding variable genes using x.low.cutoff = 0.0125, x.high.cutoff = 13, y.cutoff = -10.5. For scImpute, after comparing several combinations, we have decided to use parameters k = 11 and t = 0.5 for the case of dataset (*19*). For the dataset (*18*), we are using k=13 and t

= 0.5. MAGIC (*26*) uses its own normalization and we are using the following parameters: number of PCA dimensions = 20, k = 10 and k_a = 30 (authors recommend 3 times k).

For Figures 3G, 3H, 3I, 3J, we have calculated first 80 principal components for the dataset (*18*) and first 100 for the dataset (*19*). With ZIFA and ZIMB-WaVE, we have calculated only the first 15 components, because they simply take more than 3 days to run or give errors. In the y-axis of these figures, we are computing the mean silhouette coefficient (*27*) for each cell. The silhouette coefficient for a specific cell is given by:

$$s = \frac{b - a}{\max(a, b)}$$

where the $a$ is the mean distance between a cell and all the other cells of the same class (the class is defined by the phenotype labels provided in (*18-20*)). Parameter $b$ is the mean distance between a cell and all other cells in the next nearest cluster.

For Supplementary Figures 7, 8 and 9, we are using t-SNE representation on top of a PCA reduction, where we have selected the optimal number of principal components according to Figures 3G, 3H, 3I, 3J.

**Clustering**

The hierarchical Ward clustering method was used in Figures 4D and 5D.

**t-SNE representation**

The t-SNE representation were obtained using the default parameters, which are: Learning rate = 1000, Perplexity = 30 and Early exaggeration = 12.

**Performance**

The most computationally intensive part of our approach relates to the identification of the full set of eigenvectors and eigenvalues. We compared different off-the-shelf SVD approaches (*28*) (Supplementary Figure 15). Arpack implementation of standard SVD scales as $\mathcal{O}(N \times P \times k)$, where $N$ is the number of cells, $P$ is number of genes and $k$ is the number of dimensions. One can see that the computational complexity scales as $\mathcal{O}(N^2)$ in our case. One can take advantage of the randomized implementations of SVD (*28*) that scale as $\mathcal{O}(N \times P \times \log(k) + (N + P) \times k^2)$, provided one avoids computing all the dimensions and restricts to a small number $k$. In that case, one can scale linearly with the number of cells $\mathcal{O}(N)$. Sparse SVD does not provide additional benefits in our scenario, since we sacrifice the sparse inputs, by imposing a Z-score normalization. MP curve fitting converges in a couple of iterations on average and does not present computation burden.

**Differential expression analysis of PBMC dataset**

In Figure 4D and 4E, we are showing new potential subpopulations of PBMC cells. Here, we provide a list of the most differentiated genes for the following populations:

- Dendritic cells:
    1. Dendritic 1: CCL22, FSCN1, LAMP3, IDO1, RAMP1, DAPP1, GRP137B, CLIC2, DUSP5.
    2. Dendritic 2: FCER1A, CLEC10A, CD36, CD9, AMICA1, CTSH.
- CD14 Mono cells:
    1. Mono 1: FRP1, PDLIM7, HPSE.
    2. Mono2: CD9, TGFBI, LILRB4. These genes are associated with cells adhesion pathways.
    3. Mono3: PLA2G7, CTSL, CCL2, CXCL3, CXCL2, C5AR1, MGST1.
- Activated B-cells:
    1. B-activated 1: MIR155HG, TVP23A, NMe1, PYCR1, MRTO4, MYC, SRM, DCTPP1, EBNA1BP2, FABP5.
    2. B-activated 2: CD44 (downregulated), PRDX4, CCT2.
- Activated T-cells:
    1. T-activated 1: CD44, CCR7, GIMAP4, CD247, SELL, CLEC2D.
    2. T-activated 2: CD44 (downregulated), NR4A2, 2FAND2A, PRR7, SNGM15, CD69, CLK1.

# References Materials and Methods

## Legends of Supplementary Figures

**Supplementary Figure 1.** Distribution of spacing between the square root of consecutive Wishart matrix eigenvalues across experiments, and comparison with Wigner surmise.

**Supplementary Figure 2.** Distribution of density of eigenvalues for the Wishart matrix across experiments and comparison with Marchenko-Pastur distribution.

**Supplementary Figure 3.** Sparsity introduces changes in the distribution of the highest eigenvalue of the Wishart ensemble. This figure shows the deviations from Tracy-Widom when the fraction of nonzero values (p) goes to zero in a normal (top) and a Poisson distribution (bottom). As in the normal case, strong deviations from Tracy-Widom are observed when the Poisson parameter is small (lambda < 1).

**Supplementary Figure 4**. **A)** Calculation of Shannon Entropy for the randomized Kang et al. (*18*) dataset. This is another way of expressing the same phenomenon in Figure 2C. When the system is sparse (blue) there are eigenvectors whose entropy decreases. That is a sign of information contained in these eigenvectors. **B)** Calculation of the Inverse participation ratio (IPR) for the randomized Kang et al. (*18*) dataset. The participation ratio indicates the number of cell covariates that take part for each eigenvector. When sparsity is non-negligible in the random system (blue), there are eigenvectors which have fewer cell covariates. **C)** t-SNE representation of the randomized Kang et al. (*18*) dataset. This shows the non-linear representation of the Figure 1D. The sparsity creates privileged directions in the space as it can be seen in the right panel.

**Supplementary Figure 5. A)** Localization properties of the eigenvectors in a single-cell dataset of PBMC cells (*18*). The blue line represents the system dominated by sparsity and the red line corresponds to the system after removing sparsity. **B)** Calculation of the Shannon entropy for the eigenvectors of the PBMC (*18*) dataset. The blue (red) line corresponds to the system before (after) cleaning the sparsity. For completeness, a comparison with the non-sparse randomized dataset (green line) is plotted. **C)** Calculation of the inverse participation ratio (IPR) for the same dataset.

**Supplementary Figure 6. A)** Localization properties of the eigenvectors in a single-cell dataset of mouse cortex cells (*19*).The blue line represents the system dominated by sparsity and the red line corresponds to the system after removing sparsity. The localization in the smallest eigenvector is due to the presence of low-expressed cells (these cells could be eliminated from the analysis). **B)** Calculation of the Shannon entropy for the eigenvectors of the mouse cortex cells (*19*) dataset. The blue (red) line corresponds to the system before (after) cleaning the sparsity. For completeness, a comparison with the non-sparse

randomized dataset (green line) is plotted. **C)** Calculation of the inverse participation ratio (IPR) for the same dataset.

**Supplementary Figure 7.** Comparison of the t-SNE representation for the different methods and algorithms used in Figure 3I. This case corresponds to 15 different mouse cortex cell-phenotypes described in (*19*).

**Supplementary Figure 8.** Comparison of the t-SNE representation for the different public algorithms used in Figure 3H. This case corresponds to 7 different mouse cortex cell-phenotypes described in (*19*).

**Supplementary Figure 9.** Comparison of the t-SNE representation for the different public algorithms used in Figure 3G. This case corresponds to 13 different PBMC cell-phenotypes sequenced in (*18*) and described in (*20*).

**Supplementary Figure 10.** For the case of (*18*) dataset, comparison of some statistics before and after cleaning the sparsity **A)** Number of genes per cell. **B)** Number of transcripts per gene. **C)** Number of cells expressing a gene. **D)** Ratio of cells expressing each gene versus the average gene expression.

**Supplementary Figure 11.** For the case of (*19*) dataset, comparison of some statistics before and after cleaning the sparsity **A)** Number of genes per cell. **B)** Number of transcripts per gene. **C)** Number of cells expressing a gene. **D)** Ratio of cells expressing each gene versus the average gene expression.

**Supplementary Figure 12.** Schematic representation of the computational pipeline.

**Supplementary Figure 13:** Implementation of three steps of the algorithm for denoising of single-cell data on the example of the dataset by (*18*) **A)** Shapiro normality test vs genes sorted by the total number of transcripts. **B)** Selection of the subset of eigenvalues that correspond to the MP distribution. **C)** Sample variance of genes projected onto signal eigenvectors, eigenvectors corresponding to largest eigenvalues of MP distribution, lowest eigenvalues of MP distribution and middle part of MP distribution.

**Supplementary Figure 14.** Schematic representation of the Marchenko-Pastur distribution.

**Supplementary Figure 15.** Performance of different SVD implementations on a laptop.

**Supplementary Figure 16.** Comparison of dendritic cell populations found using RMT with populations found in paper (*29*). On the left panel, we show RMT cell-population clusters by color, the dashed rectangle focuses on the three dendritic cell populations. On the right panels, we plot the dendritic cell marker genes

identified in Villani et al. (*29*). Based on this comparison, the DC1 population from Villani et al. (*29*) corresponds to RMT Dendritic 2, whereas DC3 and DC4 are grouped in Dendritic 1. Finally, DC5/6 correspond to RMT pDC cells.

**Supplementary Figure 1**

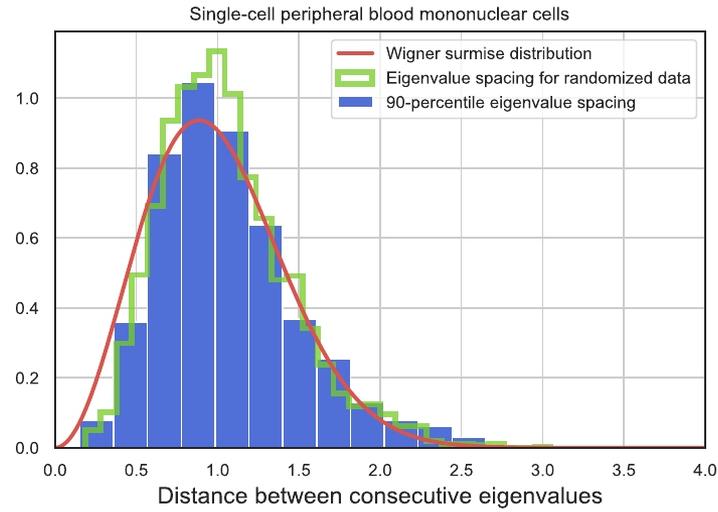
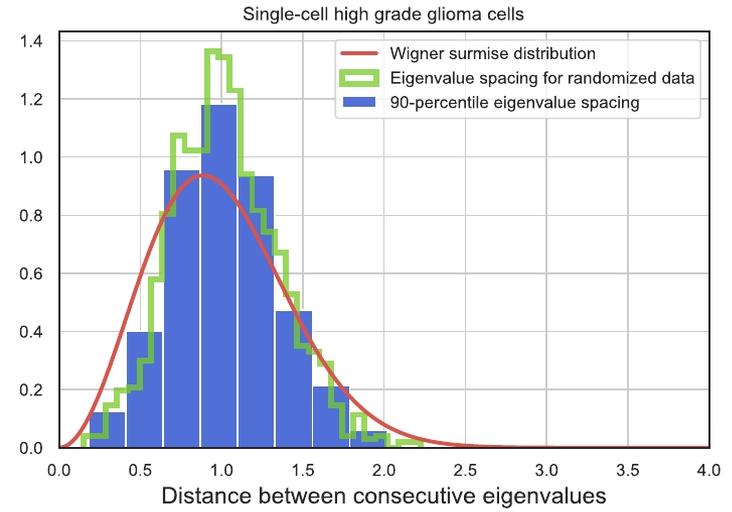
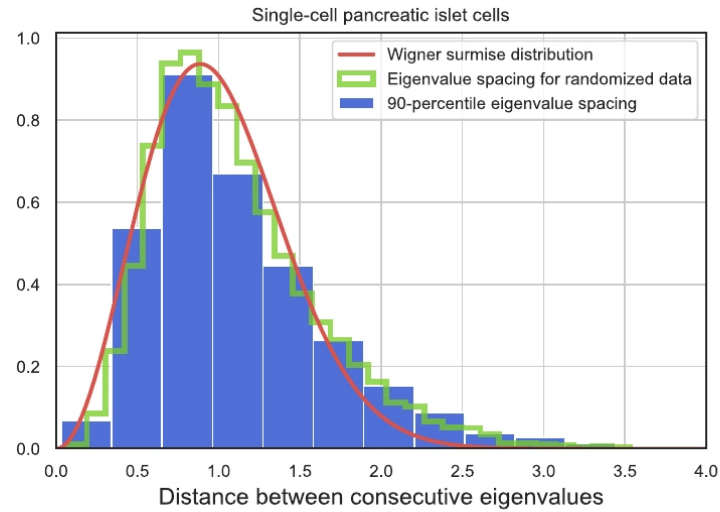
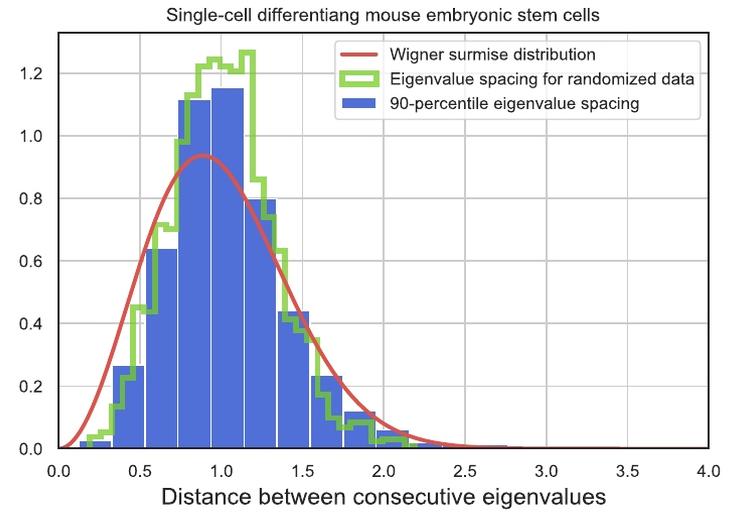

**Supplementary Figure 2**

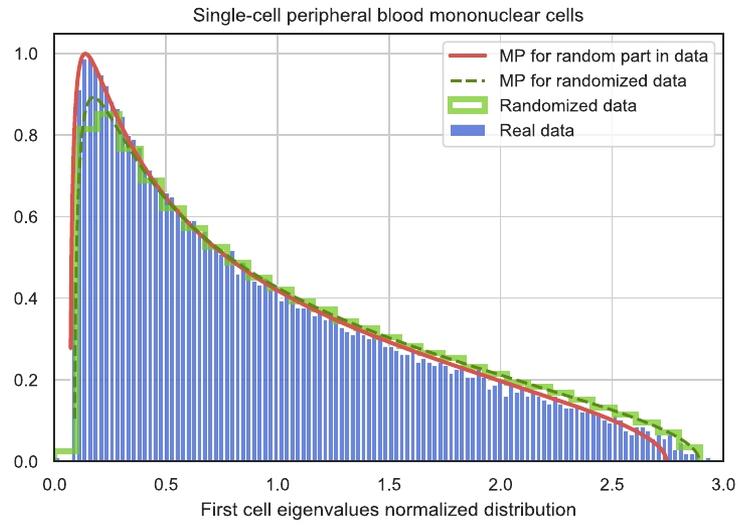
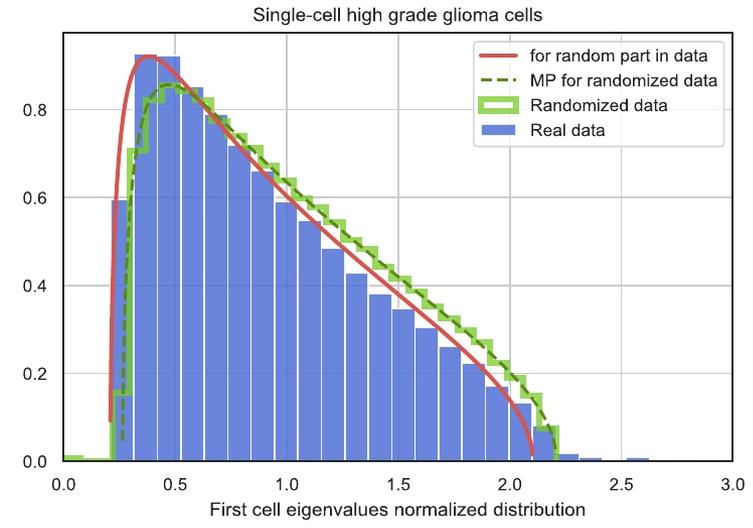
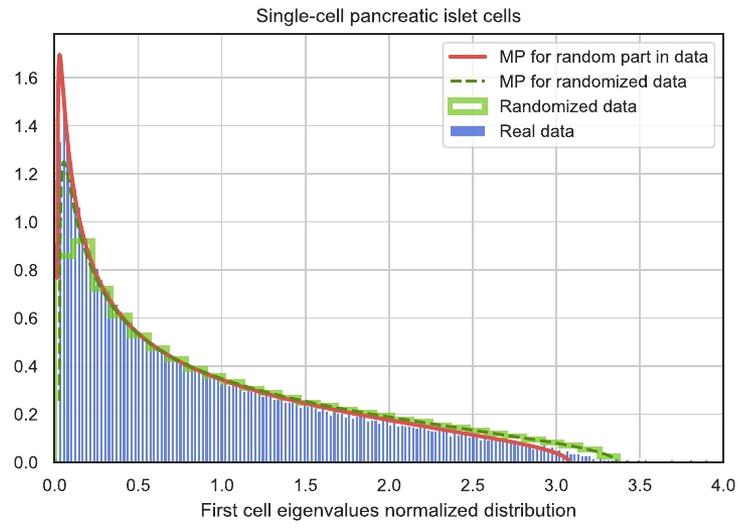
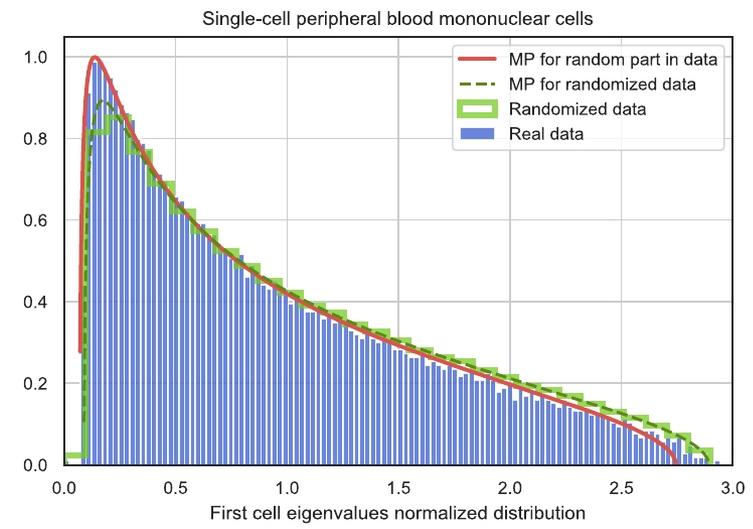

**Supplementary Figure 3**

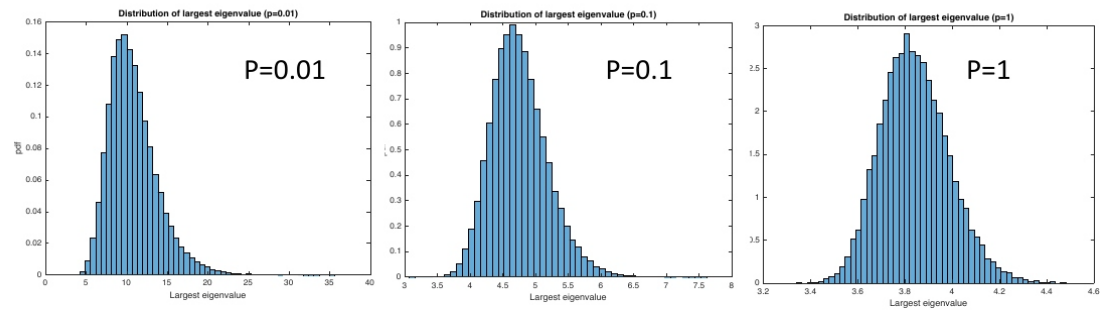

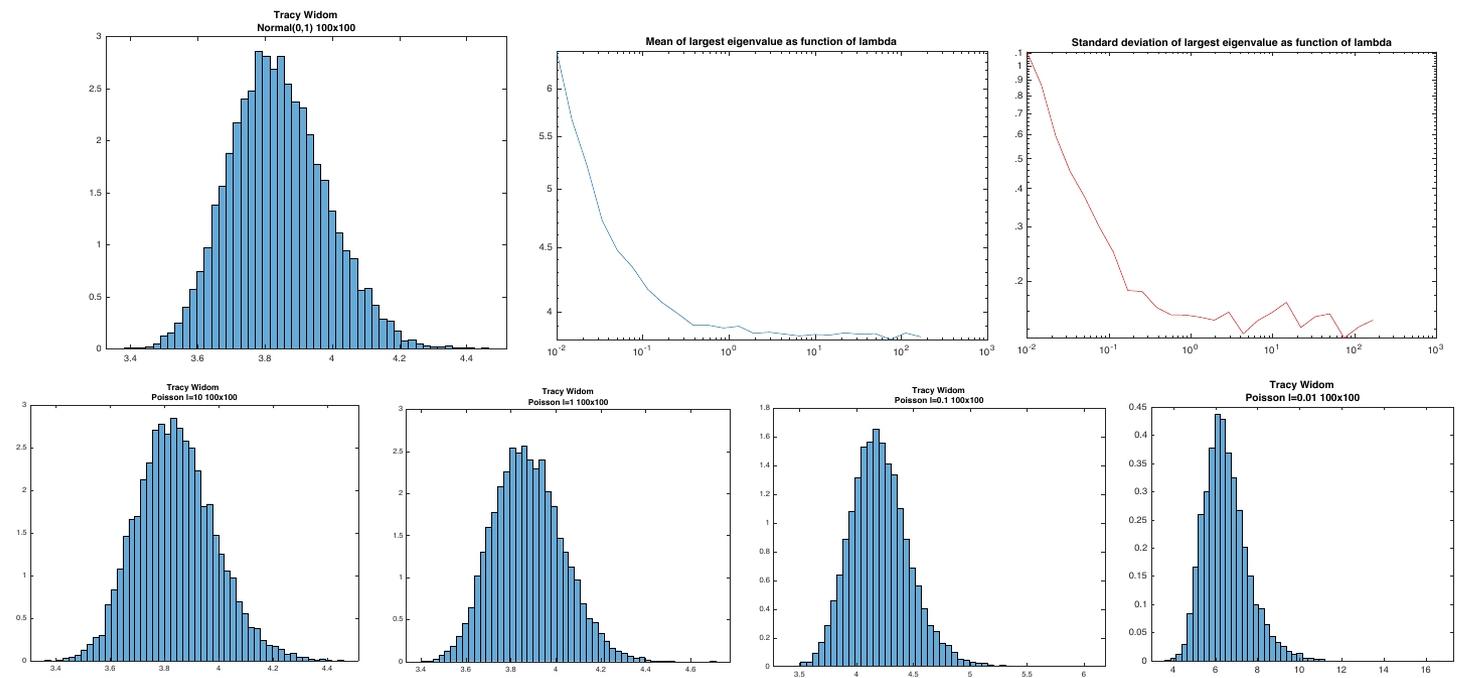

**Supplementary Fig 4**

**A**
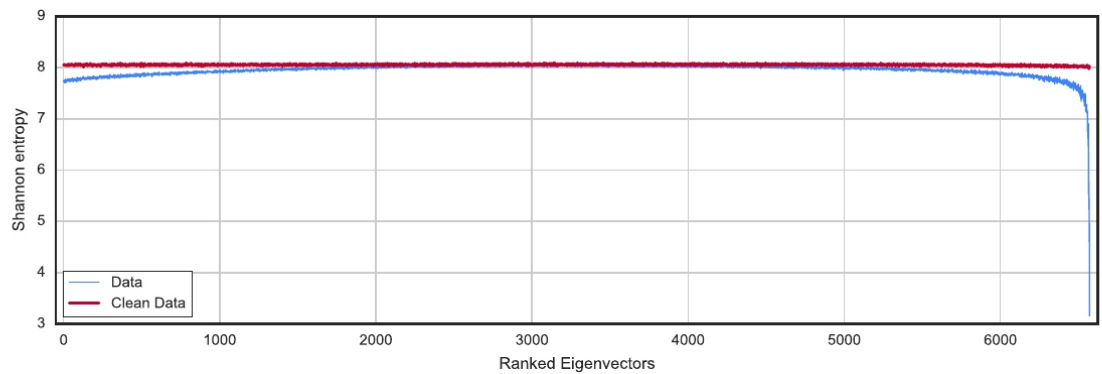

**B**
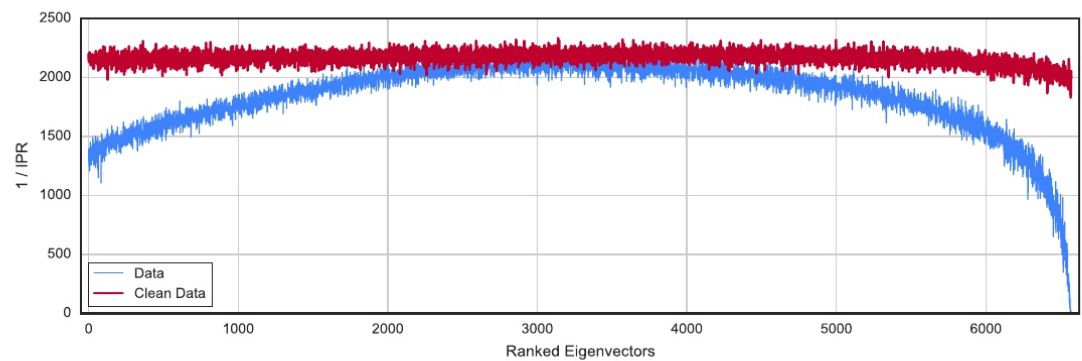

**C**
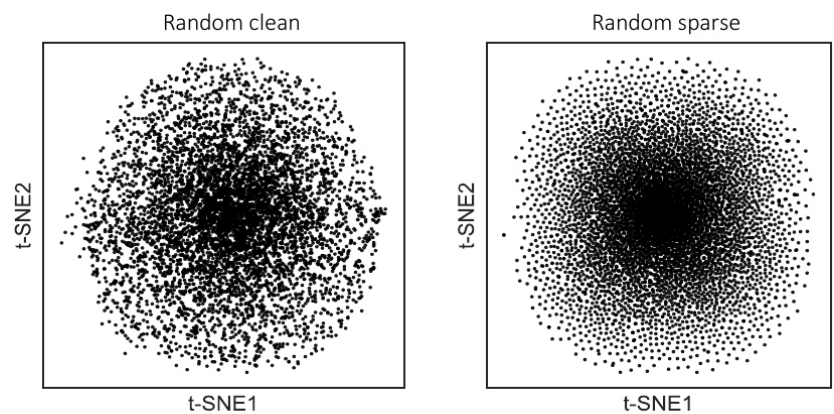

**Supplementary Fig 5**

A
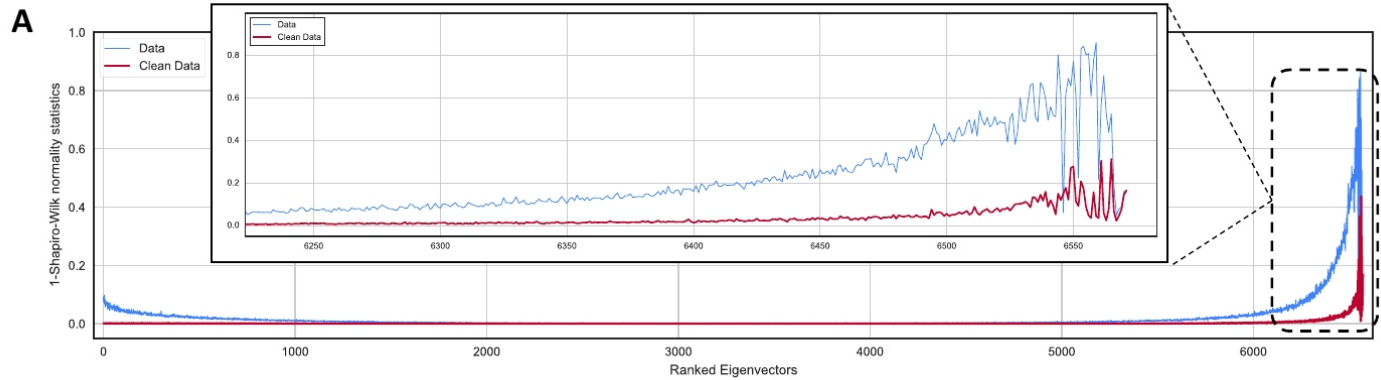

B
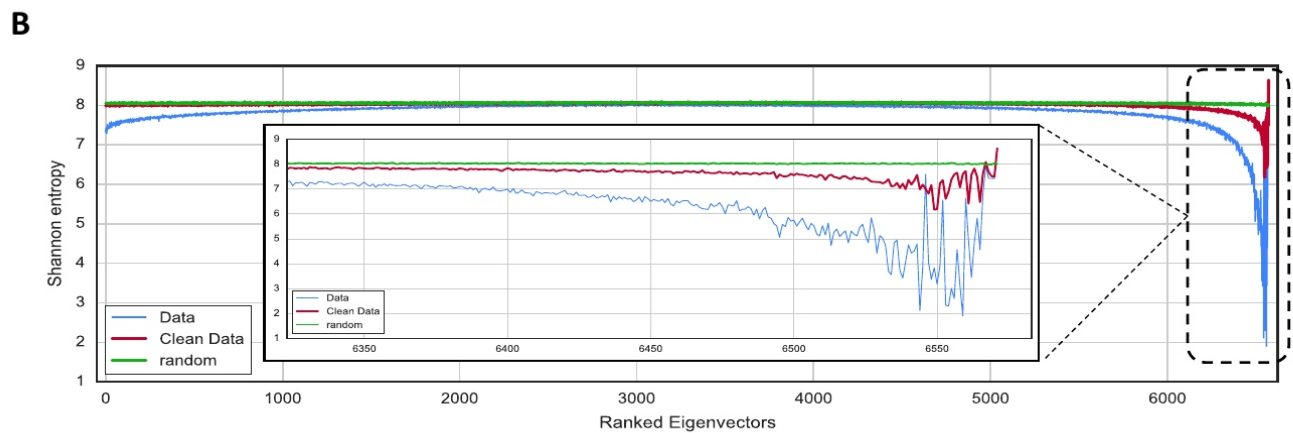

C
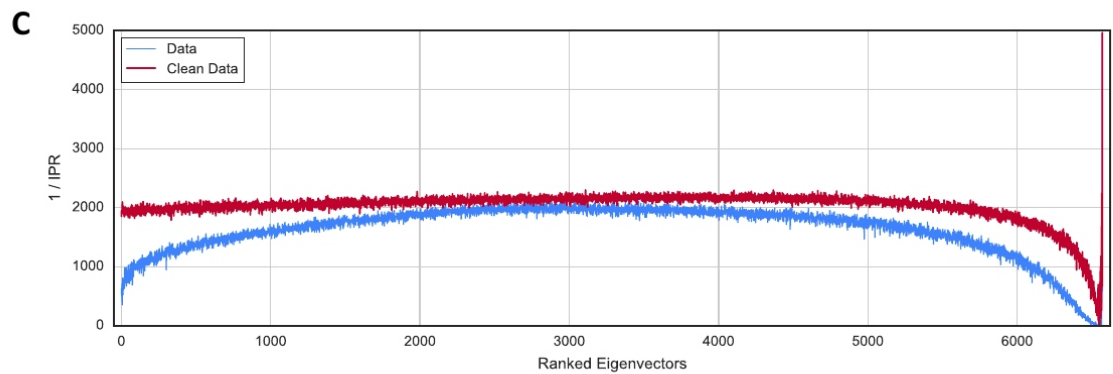

**A**

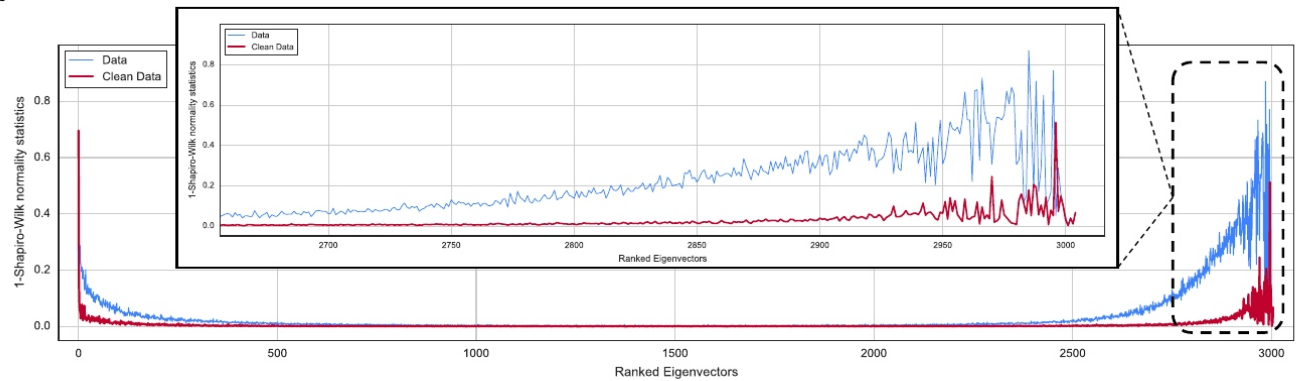

**B**

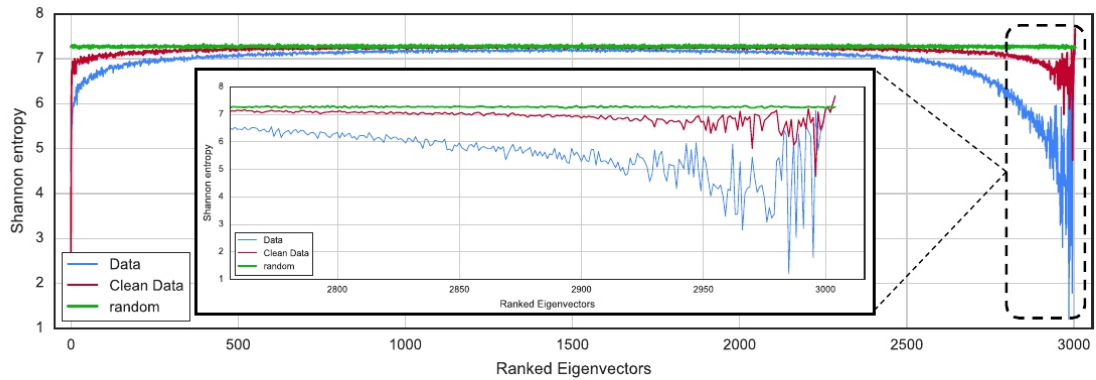

**C**

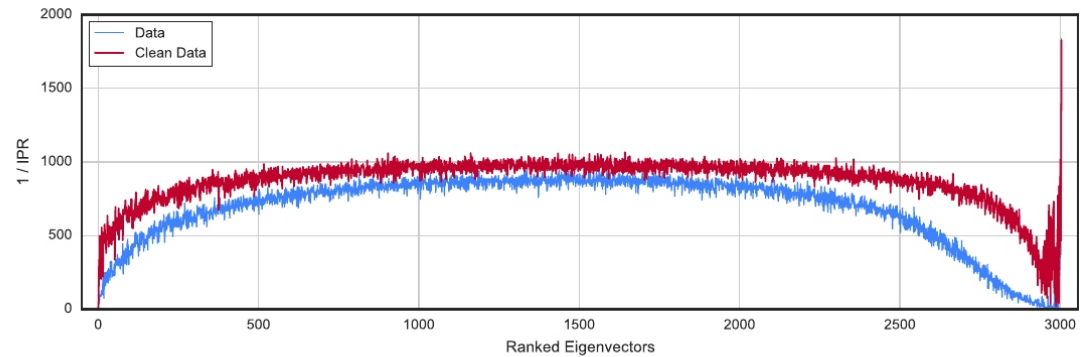

**Supplementary Figure 7**

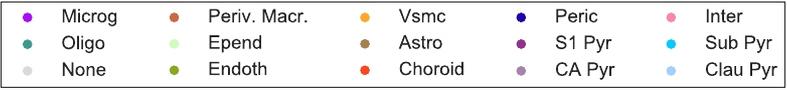

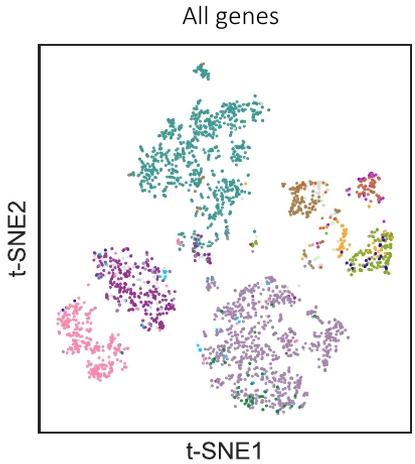
All genes

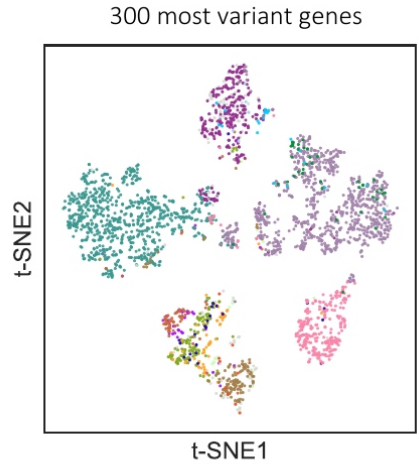
300 most variant genes

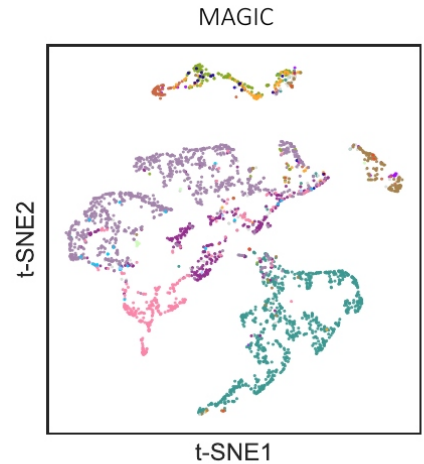
MAGIC

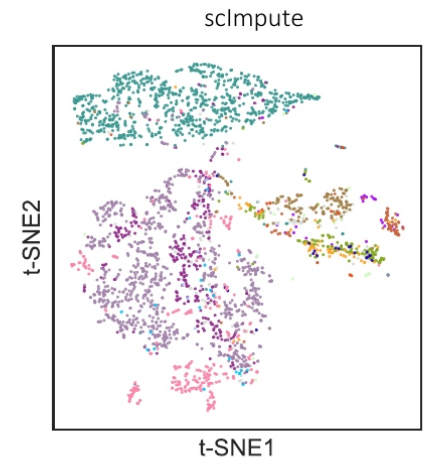
scImpute

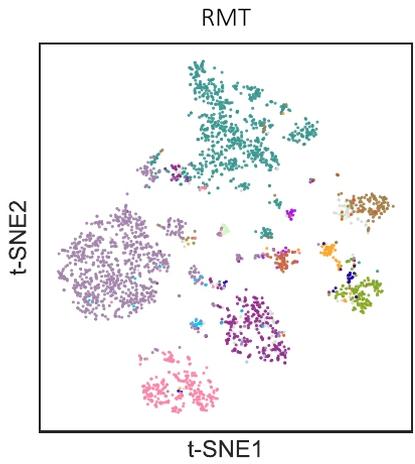
RMT

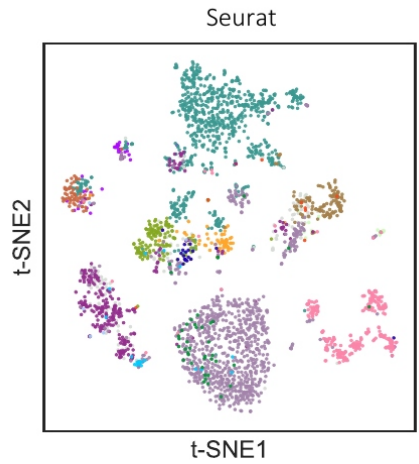
Seurat

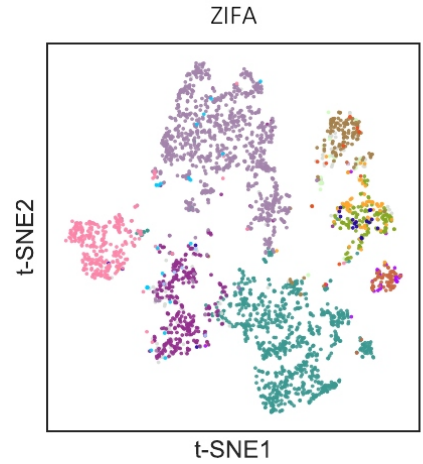
ZIFA

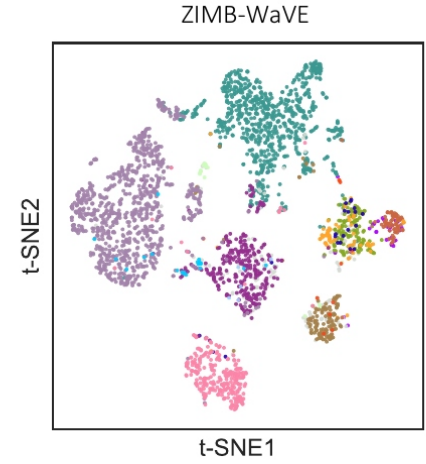
ZIMB-WaVE

**Supplementary Figure 8**

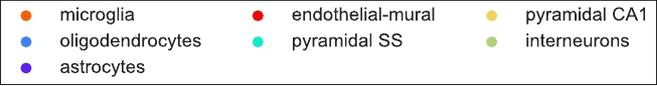
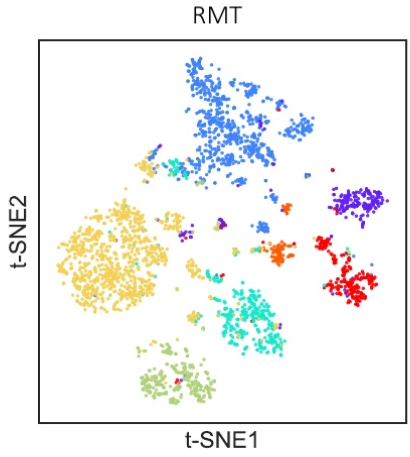
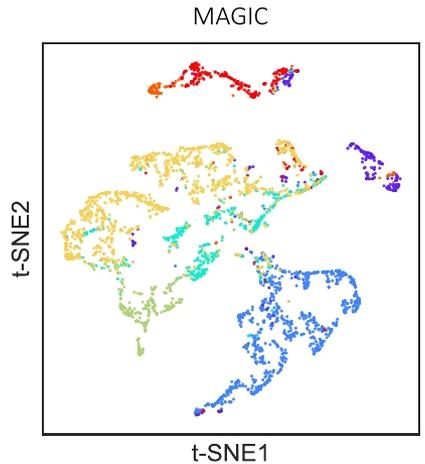
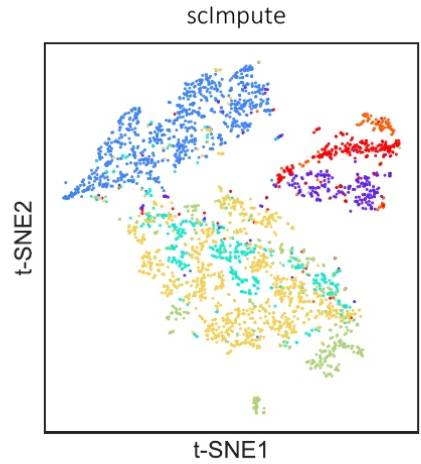
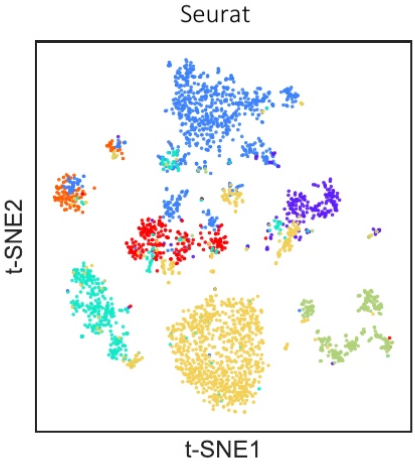
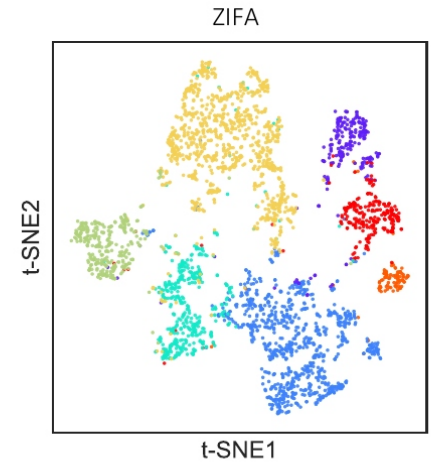
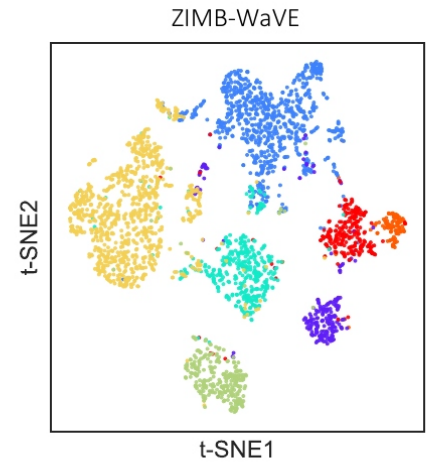

**Supplementary Figure 9**

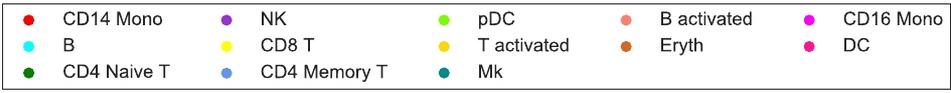

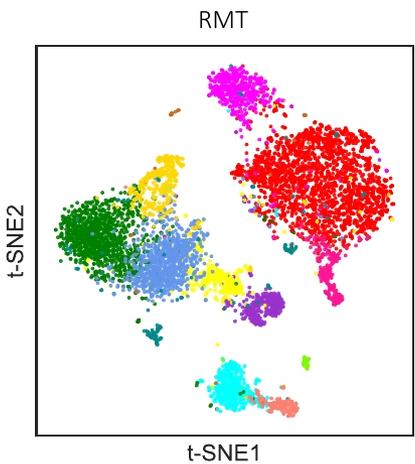
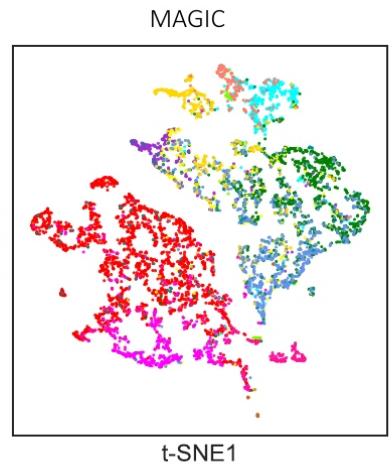
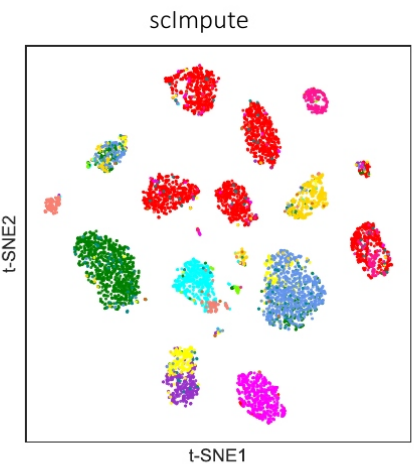
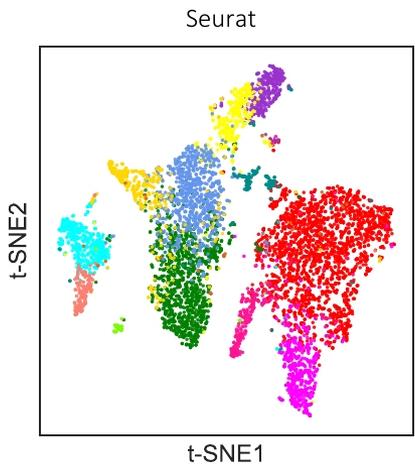
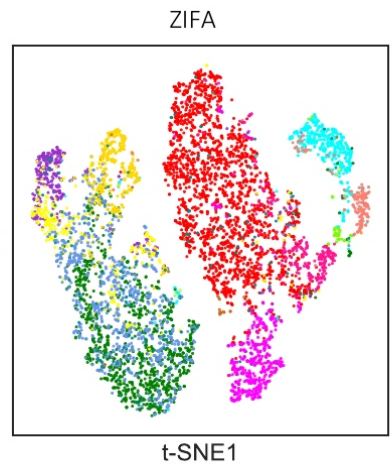
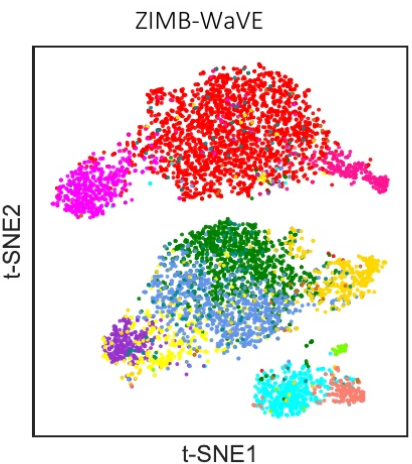

**Supplementary Figure 10**

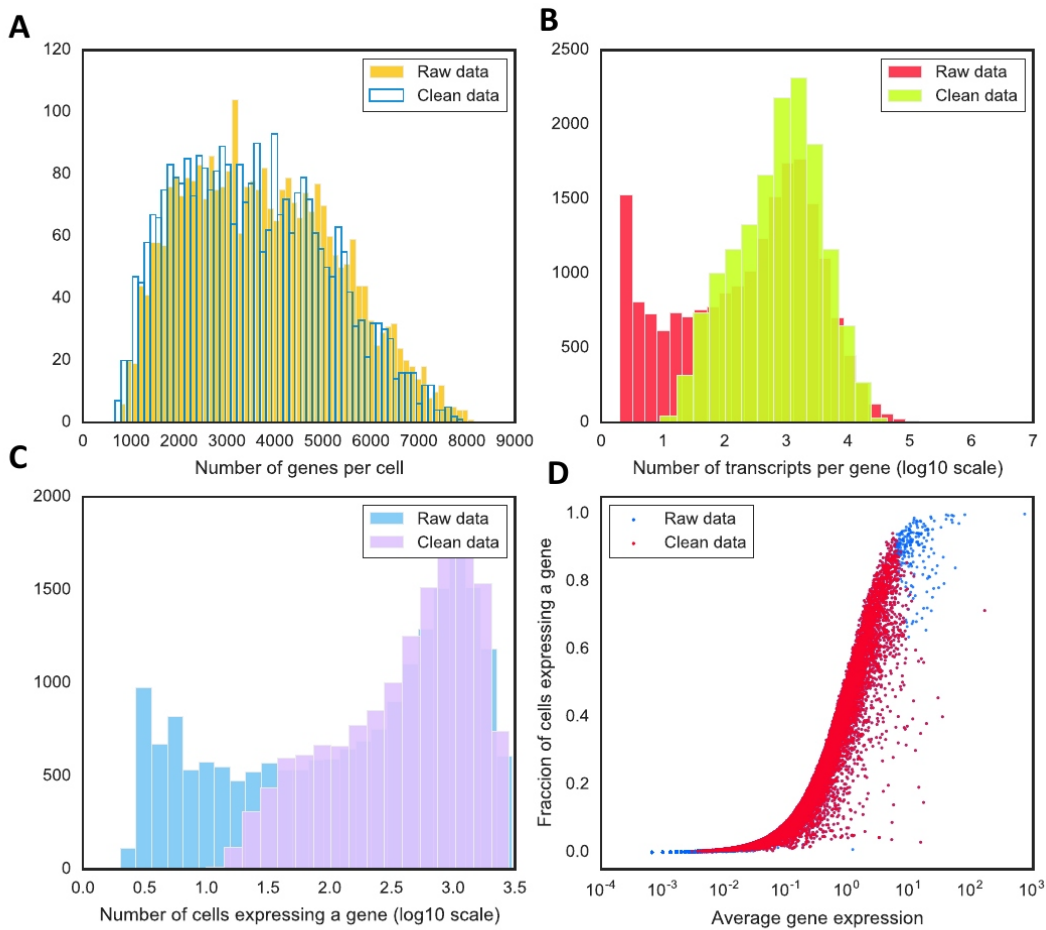

**Supplementary Figure 11**

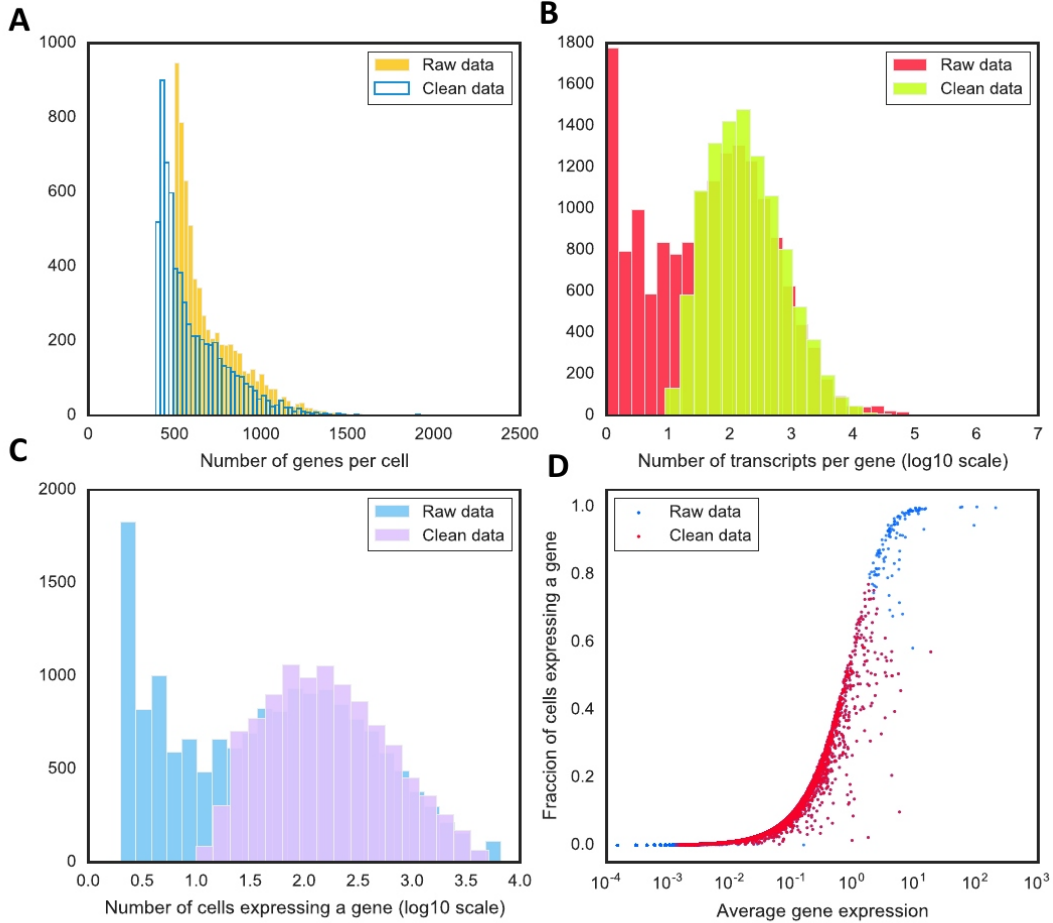

**Supplementary Figure 12**

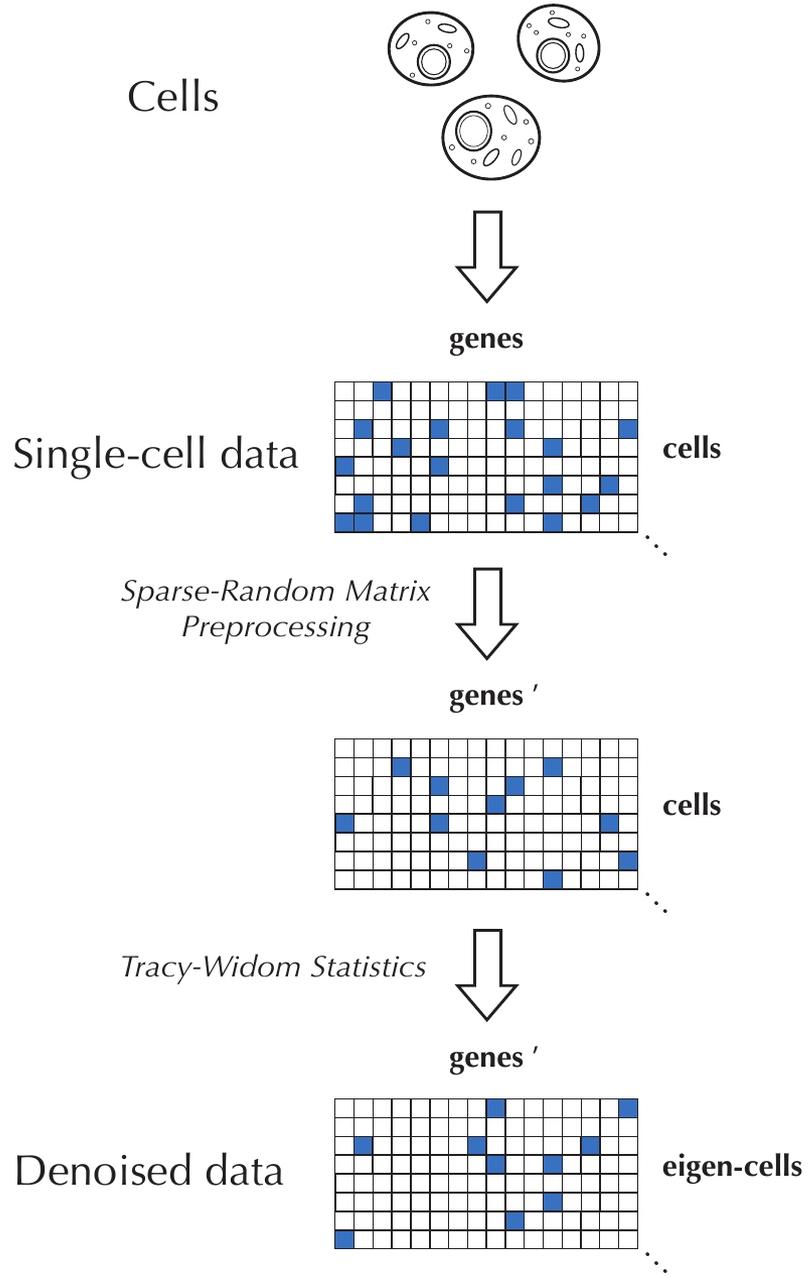

# Supplementary Figure 13

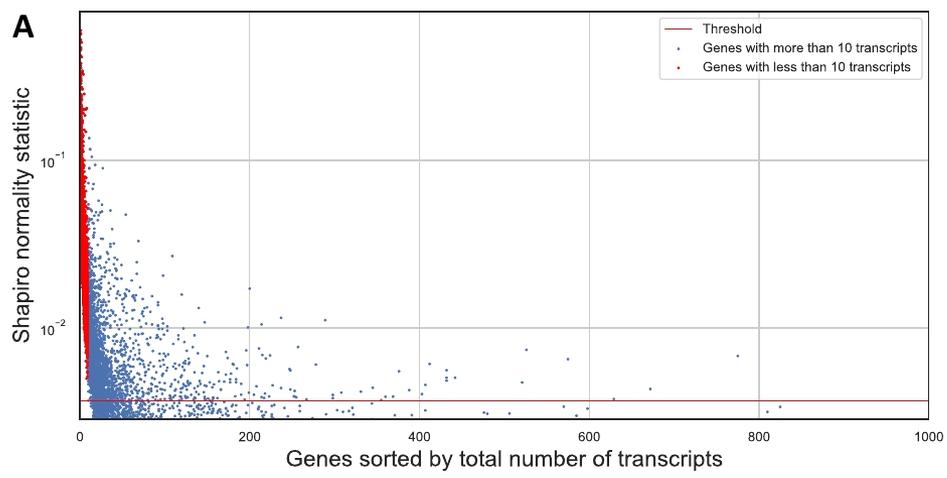
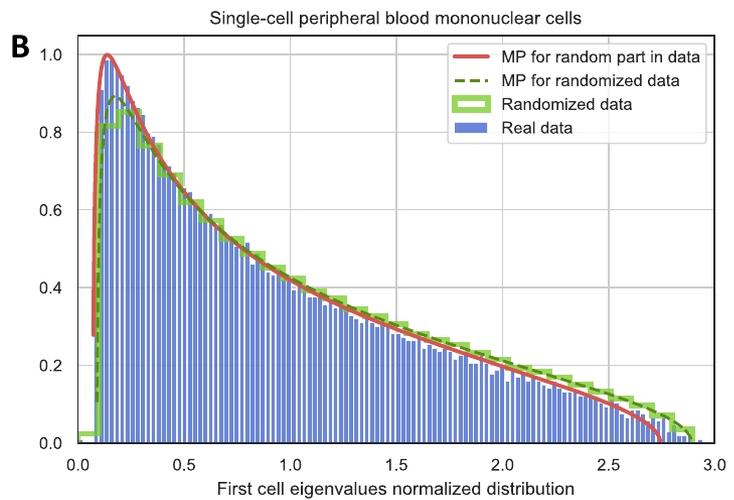
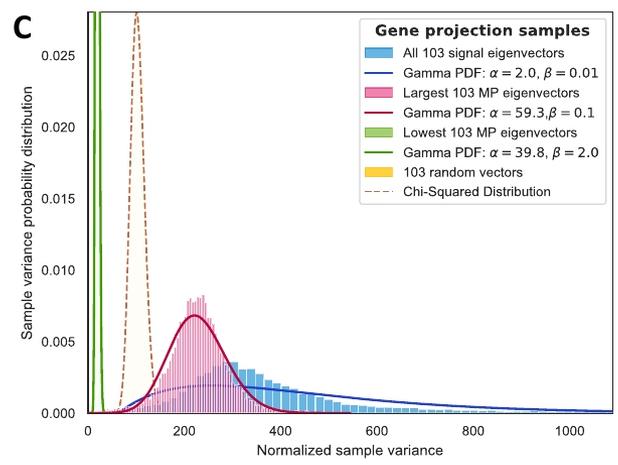
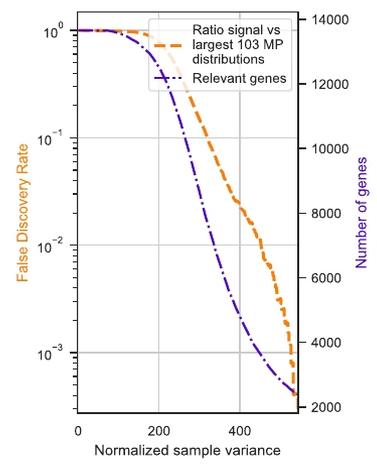

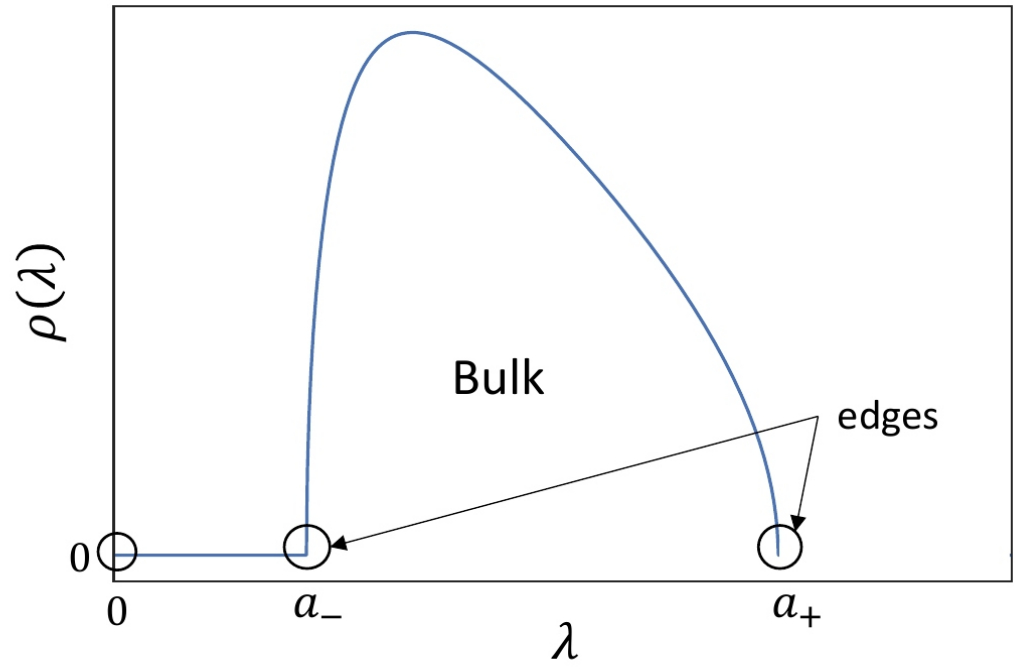

**Supplementary Figure 15**

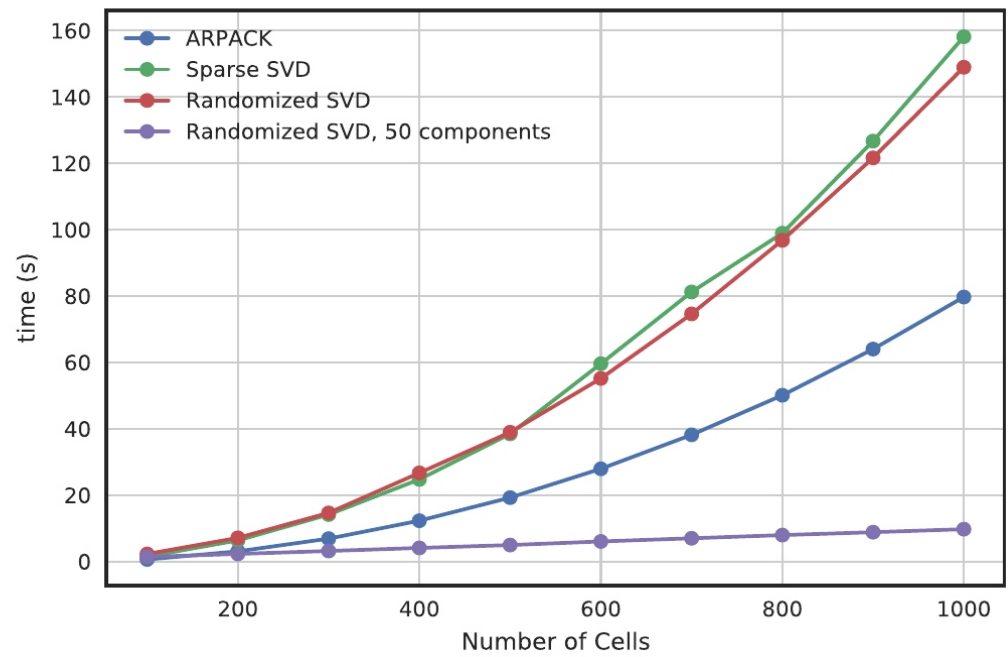

**Supplementary Figure 16**

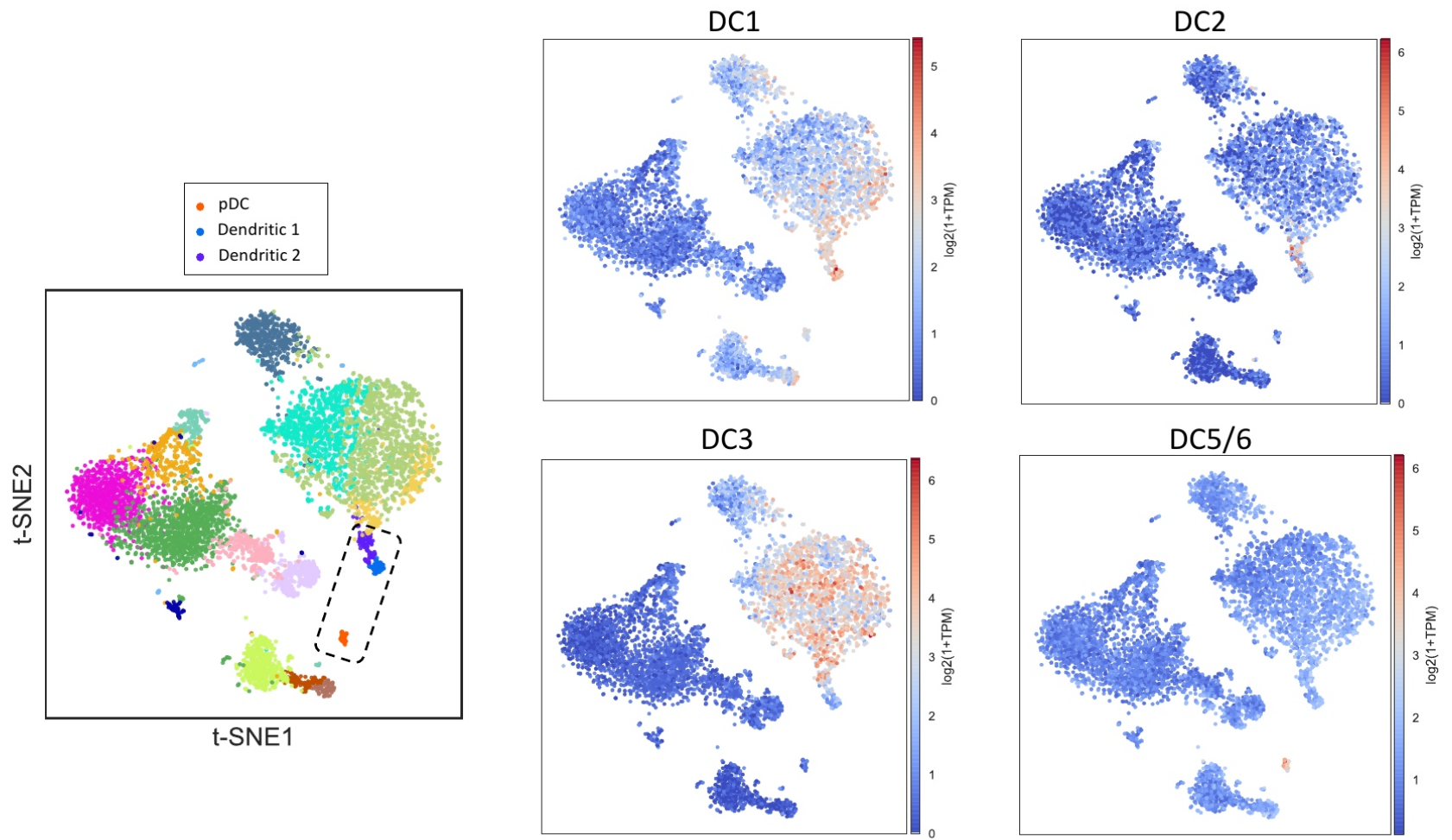